\documentclass[journal]{IEEEtran}
\usepackage{ifpdf}

\usepackage{cite}

\ifCLASSINFOpdf
  \usepackage[pdftex]{graphicx}
\else
  \usepackage[dvips]{graphicx}
\fi
\usepackage{amsmath,amssymb}
\usepackage{dblfloatfix}

\usepackage{color}
\usepackage{threeparttable}

\newcounter{MYtempeqncnt}

\begin{document}
%
\title{Arbitrary Power-Conserving Field Transformations with Passive Lossless Omega-Type \\ Bianisotropic Metasurfaces}
%
%
%

\author{Ariel~Epstein,~\IEEEmembership{Member,~IEEE,}
        and~George~V.~Eleftheriades,~\IEEEmembership{Fellow,~IEEE}
\thanks{The authors are with the Edward S. Rogers Sr. Department of Electrical and
Computer Engineering, University of Toronto, Toronto, ON M5S 2E4 Canada (e-mail: ariel.epstein@utoronto.ca; gelefth@waves.utoronto.ca).}
\thanks{Manuscript received April 1, 2016; revised June 24, 2016.}}

%
%

\markboth{IEEE Transactions on Antennas and Propagation,~Vol.~64, No.~9, September~2016}%
{Epstein and Eleftheriades: Arbitrary field transformations with OBMSs}
%


\IEEEspecialpapernotice{(DOI: 10.1109/TAP.2016.2588495) ~\copyright~2016 IEEE}

\maketitle

\begin{abstract}
We present a general theory for designing realistic omega-type bianisotropic metasurfaces (O-BMSs), unlocking their full potential for molding electromagnetic fields. These metasurfaces, characterized by electric surface impedance, magnetic surface admittance, and magnetoelectric coupling coefficient, were previously considered for wavefront manipulation. However, previous reports mainly considered plane-wave excitations, and implementations included cumbersome metallic features. In this work, we prove that \emph{any} field transformation which locally conserves real power can be implemented via passive and lossless meta-atoms characterized by closed-form expressions; this allows rigorous incorporation of arbitrary source and scattering configurations. Subsequently, we show that O-BMS meta-atoms can be implemented using an asymmetric stack of three impedance sheets, an appealing structure for printed circuit board fabrication. Our formulation reveals that, as opposed to Huygens' metasurfaces (HMSs), which exhibit negligible magnetoelectric coupling, O-BMSs are not limited to controlling the phase of transmitted fields, but can rather achieve high level of control over the amplitude and phase of reflected fields. This is demonstrated by designing O-BMSs for reflectionless wide-angle refraction, independent surface-wave guiding, and a highly-directive low-profile antenna, verified with full-wave simulations. This straightforward methodology facilitates development of O-BMS-based devices for controlling the near and far fields of arbitrary sources in complex scattering configurations. 
\end{abstract}


%
\IEEEpeerreviewmaketitle

\section{Introduction}
%
%
%
%
\label{sec:introduction}
\IEEEPARstart{M}{etasurfaces} have attracted considerable attention lately due to their demonstrated ability to control numerous diverse features of electromagnetic fields \cite{Holloway2012,Kildishev2013,Yu2014}. These structures consist of subwavelength polarizable particles (meta-atoms) arranged on a plane. This forms a low profile device of subwavelength thickness, which can be modelled by equivalent boundary conditions. 

If the the meta-atoms exhibit dominant electric polarizability, e.g. as for metallic wire arrays, their response can be described by an electric surface impedance $Z_{se}$; such metasurfaces were used, for instance, for tailoring the phase profile of an incident wavefront, albeit with relatively low efficiency \cite{Yu2011}. If the the meta-atoms exhibit dominant magnetic polarizability, e.g. when metallic patch arrays are backed by a perfect electric conductor (PEC), the metasurface response can be described by a magnetic surface admittance $Y_{sm}$; these types of metasurfaces are especially useful for guiding surface waves or leaky waves \cite{Maci2011,Minatti2015}, or for reflector applications \cite{Estakhri2014}. 

It was recently demonstrated that by utilizing collocated electric and magnetic polarizable meta-atoms, phase and polarization of transmitted fields can be controlled with high transmission coefficients \cite{Pfeiffer2013, Monticone2013, Selvanayagam2013}. These so-called Huygens' metasurfaces (HMSs) can be formed by properly aligning metallic wires and loops \cite{Wong2014,Wong2015}, or, equivalently, utilizing \emph{symmetric} 3-layer stacks of electric impedance sheets \cite{Pfeiffer2014,Pfeiffer2014_2}; in optical frequencies, dielectric particles exhibiting simultaneous electric and magnetic resonances can be used as Huygens' meta-atoms \cite{Decker2015,Shalaev2015,Campione2015,Yu2015}. Following the equivalence principle, the ability to effectively induce electric and magnetic currents on a plane allows, in principle, to impose arbitrary field discontinuities; this, in turn, would allow the implementation of arbitrary field transformations.

Nevertheless, it turns out that transforming a given incident field to a desirable transmitted field generally requires electric surface impedance and magnetic surface admittance which include meta-atoms implementing local loss and gain \cite{Epstein2015_3}. As this is typically undesirable from an implementation perspective, numerical optimization schemes may sometimes be used to refine these field quantities to reduce the required loss and gain values \cite{Pfeiffer2013,Achouri2015}. Alternatively, in previous work we have shown that if the reflected and transmitted fields are semianalytically stipulated such that local impedance equalization and local power conservation are satisfied (at each point along the metasurface), passive and lossless HMS designs can be guaranteed for a given (arbitrary) source field \cite{Selvanayagam2013,Epstein2014}. While this general design scheme can be harnessed to devise exceptional electromagnetic radiators, for instance \cite{Epstein2016}, it does not allow control of the reflected fields, which is desirable many times.

In contrast, omega-type bianisotropic metasurfaces (O-BMSs) have been shown to allow independent control of both reflection and transmission magnitude and phase of normally-incident plane waves \cite{Radi2014_1, Asadchy2015}. While in HMSs applied electric fields induce electric currents (related to $Z_{se}$) and applied magnetic fields induce (equivalent) magnetic currents (related to $Y_{sm}$) \cite{Epstein2015_3}, O-BMSs feature also significant magnetoelectric coupling $K_{em}$ \cite{Radi2013,Tretyakov2015}. This means that applied electric fields induce \emph{also} (equivalent) magnetic currents, and applied magnetic fields induce \emph{also} electric currents on the surface. This additional degree of freedom allows engineering the field interaction with each facet of the metasurface separately, without violating the requirement for passive and lossless implementation. 

Although the metasurfaces reported in \cite{Radi2014_1, Asadchy2015} demonstrate both theoretically and experimentally the ability of O-BMSs to independently control their forward and backward scattering properties, two major issues still prevent this concept from revealing its full potential. First, the formalism developed in \cite{Radi2014_1, Asadchy2015} only applies to normally-incident plane waves, providing tools to engineer the magnitude and phase of reflected and transmitted waves. While this may be suitable for approximated ray-optical scenarios \cite{Asadchy2015}, such a formalism is not capable of accommodating arbitrary source configurations, e.g. including localized sources and scatterers. In these types of scenarios, usually encountered in practical applications such as antennas, light-emitting devices, waveguides, and cavities, a scheme to engineer the metasurface interaction with rich spatial spectra (including obliquely-propagating and evanescent components) has to be developed. As has been extensively discussed in \cite{Epstein2014, Epstein2014_2, Epstein2015_3, Estakhri2016}, relying on phase-shift stipulation design schemes for general electromagnetic field manipulation is inaccurate, since these schemes disregard the magnitude variations which must accompany phase variations. 

Second, the physical structure of the bianisotropic meta-atoms presented in \cite{Radi2014_1, Asadchy2015} (and also in \cite{Asadchy2015_1}) is quite cumbersome and has to be manually-fabricated, which makes it hard to realize such O-BMSs in practice. For practical applications, it would be useful to devise a realization of O-BMSs which is compatible with standard fabrication techniques. For example, for most metasurfaces operating at microwave frequencies, a physical design which can be fabricated in common printed circuit board (PCB) manufacturing facilities has been reported to date (e.g., \cite{Pfeiffer2013, Pfeiffer2014_1, Pfeiffer2014_3, Wong2014, Wong2015, Minatti2015}).

In this paper, we present solutions to both these problems. First, we rigorously derive closed-form expressions which prescribe the required (macroscopic) O-BMS properties to implement a given (arbitrary) transformation between two sets of electromagnetic fields (Subsection \ref{subsec:macro_design}). We show that if the real power crossing the metasurface is locally-conserved at each point on the surface, the O-BMS can be implemented using passive and lossless meta-atoms. This facilitates straightforward design of realistic O-BMSs, capable of interacting with complex fields and source configurations to perform a myriad of electromagnetic manipulations. Second, using an impedance-matrix formalism analogous to \cite{Selvanayagam2013_2} we show that a general omega-type bianisotropic meta-atom can be realized by \emph{asymmetric} three-layer stack of electric impedance sheets (a variation on  \cite{Pfeiffer2013_2,Monticone2013}; Subsection \ref{subsec:micro_design}). We utilize this model, which is compatible, in principle, with standard PCB fabrication techniques, to define and perform full-wave simulations of O-BMSs using commercially-available finite-element solver (ANSYS HFSS). It should be noted that in contrast to the theory developed in \cite{Pfeiffer2014_3}, showing that stacking of a number impedance sheet tensors allows implementation of a general scattering matrix response, the microscopic design procedure formulated herein utilizes a specific 3-layer asymmetric scalar sheet impedance stack to implement omega-type meta-atoms, derives analytical formulas for the sheet impedance values, and points out the close relation to Huygens' meta-atoms.

Based on these developments, we demonstrate the wide variety of functionalities that can be achieved with PCB-compatible passive and lossless O-BMSs. We first design an O-BMS to implement reflectionless wide-angle plane-wave refraction (Subsection \ref{subsec:refraction}). As \emph{all} the incident power is transmitted through the O-BMS, local power conservation is satisfied, facilitating the use of our methodology. This demonstrates the ability of O-BMSs to support extreme and abrupt change in the wave impedance without incurring reflections nor active nor lossy components; such a performance is not achievable with standard refracting HMSs \cite{Epstein2014_2}. Next, we design an O-BMS that supports independent surface wave propagation on both its facets (Subsection \ref{subsec:surface_wave}); the surface wave propagation and decay constants can be arbitrarily chosen. This example demonstrates the possibility to incorporate guided and evanescent modes into the design of O-BMS-based devices. In this case, \emph{no} real power is crossing the metasurface, allowing us to use the proposed design procedure. Lastly, we harness the derived formalism to individually control the reflection coefficients of different cavity modes, excited by a single localized source (Subsection \ref{subsec:cavity_excited_antenna}); in this case, the O-BMS acts as an advanced \emph{partially}-reflecting surface (PRS). Utilizing the degrees of freedom provided by this rather complex source configuration, we show that we can guarantee that only a \emph{single} cavity mode is excited near the metasurface, leading to a highly-structured and uniform illumination of the O-BMS (arbitrarily-large) aperture. Using the O-BMS capability to control the fields on its top facet as well, a uniform phase is established on the aperture, resulting in a low-profile cavity-excited antenna exhibiting $100\%$ aperture illumination efficiency. This device outperforms our previously-reported cavity-excited HMS antenna \cite{Epstein2016}, while requiring only half the device thickness.

These diverse examples provide a glance into the great potential of O-BMSs for a myriad of applications, made accessible thanks to the formalism developed herein. They indicate that O-BMSs designed following our methodology can reliably control reflected and transmitted fields even in highly-intricate configurations, including complex source-fields and scatterers. At the same time, once the fields above and below the metasurface are stipulated and local power conservation is verified, the design procedure becomes straightforward, and is guaranteed to yield appealing passive and lossless implementation requirements.

\section{Theory}
\label{sec:theory}
\subsection{Macroscopic (metasurface) design}
\label{subsec:macro_design}
\begin{figure}[!t]
\centering
\includegraphics[width=8cm]{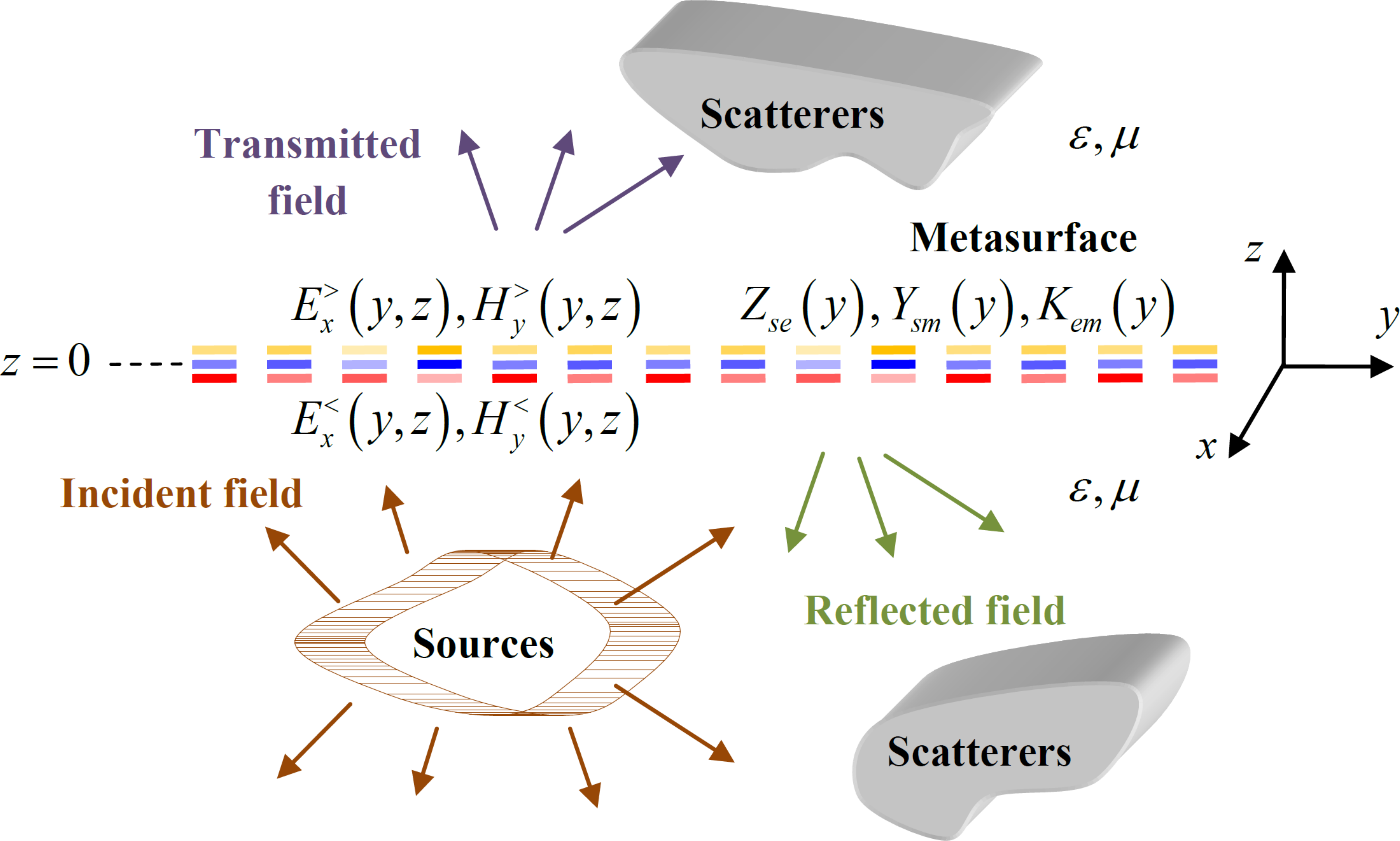}%
\caption{Physical configuration of an omega bianisotropic metasurface implementing the transformation between $\vec{E}^<\left(y,z\right)$,$\vec{H}^<\left(y,z\right)$ and $\vec{E}^>\left(y,z\right)$,$\vec{H}^>\left(y,z\right)$. Although the schematic depicts a scenario where the sources are limited to the lower half space, in the general case sources and scatterers can be distributed everywhere in space, as long as the stipulated fields satisfy the relevant boundary, continuity, source, and radiation conditions.}
\label{fig:physical_configuration}
\end{figure}

For simplicity, we derive a design methodology suitable for 2D configurations ($\partial/\partial x =0$) excited transverse electric (TE) fields ($E_z=E_y=H_x=0$). 
We consider, thus, an O-BMS situated at $z=0$ embedded in \textcolor{black}{a} homogeneous medium with permittivity $\epsilon$ and permeability $\mu$. The half-spaces below ($z<0$) and above ($z>0$) the metasurface may contain any time-harmonic sources $e^{j\omega t}$ and any scattering geometries, as depicted in Fig. \ref{fig:physical_configuration}. The wavenumber and wave impedance in the unbounded medium are given by $k=\omega\sqrt{\mu\epsilon}$ and $\eta=\sqrt{\mu/\epsilon}$, respectively.

We recall that our aim is to design a passive and lossless O-BMS implementing a desirable transformation between the fields below the metasurface ($z<0$) $\left\{E_x^<\left(y,z\right),H_y^<\left(y,z\right),H_z^<\left(y,z\right)\right\}$ and above it ($z>0$) $\left\{E_x^>\left(y,z\right),H_y^>\left(y,z\right),H_z^>\left(y,z\right)\right\}$. Therefore, the input to our design procedure is these two sets of electromagnetic fields. These fields can be arbitrarily stipulated, as long as they meet two conditions: first, they must satisfy Maxwell's equations (including the relevant boundary conditions, source conditions, and radiation conditions \cite{FelsenMarcuvitz1973}) at each one of the half-spaces ($z\neq0$); second, the fields at the metasurface facets $z\rightarrow 0^{\pm}$ should \emph{locally}-conserve the real power crossing the metasurface. Explicitly, this latter \emph{local} power conservation condition reads \cite{Epstein2014}
\begin{align}
P_z^{-}\left(y\right)&=\frac{1}{2}\Re\left\{E_x^-\left(y\right)H_y^{-*}\left(y\right)\right\} \nonumber \\
	&=\frac{1}{2}\Re\left\{E_x^+\left(y\right)H_y^{+*}\left(y\right)\right\}=P_z^{+}\left(y\right),
\label{equ:local_power_conservation}
\end{align}
where $E_x^+\left(y\right)\triangleq\left.E_x^>\left(y,z\right)\right|_{z\rightarrow0^+}$ and $E_x^-\left(y\right)\triangleq\left.E_x^<\left(y,z\right)\right|_{z\rightarrow0^-}$ 
, respectively, are the tangential electric fields just above and below the metasurface (and analogously for the tangential magnetic fields $H_y$), and $P_z^{\pm}\left(y\right)$ is the real part of the $z$-directed component of the Poynting vector.

As discussed in Section \ref{sec:introduction}, in bianisotropic metasurfaces, applied electric and magnetic fields induce both electric ($\vec{J}_s$) and magnetic ($\vec{M}_s$) surface currents \cite{Tretyakov2015}. Therefore, for scalar omega-type BMSs, the relation between the average applied tangential fields  and the induced currents can be written as \cite{Radi2013}
\begin{equation}
\left\{\!\!\!
	\begin{array}{l}
	\vspace{3pt}
	\vec{E}_{t,\mathrm{avg}}\left(y\right)\!=\!Z_{se}\left(y\right)\vec{J}_s\left(y\right)
		\!-\!K_{em}\left(y\right)\left[\hat{z}\times\vec{M}_s\left(y\right)\right] \\	
	\vec{H}_{t,\mathrm{avg}}\left(y\right)\!=\!Y_{sm}\left(y\right)\vec{M}_s\left(y\right)
		\!-\!K_{em}\left(y\right)\left[\hat{z}\times\vec{J}_s\left(y\right)\right]\!\!,
	\end{array}
\right.
\label{equ:average_fields_induced_currents}
\end{equation}
where $Z_{se}\left(y\right)$, $Y_{sm}\left(y\right)$, and $K_{em}\left(y\right)$, respectively, are the surface electric impedance, surface magnetic admittance, and magnetoelectric coupling coefficient at the point $y$ on the O-BMS. The average tangential electric and magnetic fields are defined as
$\vec{E}_{t,\mathrm{avg}}\left(y\right)\triangleq\frac{1}{2}\left[\vec{E}_{t}^+\left(y\right)+\vec{E}_{t}^-\left(y\right)\right]$ and $\vec{H}_{t,\mathrm{avg}}\left(y\right)\triangleq\frac{1}{2}\left[\vec{H}_{t}^+\left(y\right)+\vec{H}_{t}^-\left(y\right)\right]$, respectively, $\vec{E}_{t}^\pm\left(y\right)$ and $\vec{H}_{t}^\pm\left(y\right)$ being the tangential fields at $z\rightarrow0^{\pm}$.

On the other hand, the induced surface currents introduce a discontinuity in the tangential fields following the relations \cite{Pfeiffer2013,Balanis1997}
\begin{equation}
\left\{\!\!\!
	\begin{array}{l}
	\vspace{3pt}
	\vec{J}_s\left(y\right)=\hat{z}\times\Delta\vec{H}_{t}\left(y\right) \\
	\vec{M}_s\left(y\right)=-\hat{z}\times\Delta\vec{E}_{t}\left(y\right),
	\end{array}
\right.
\label{equ:field_difference_induced_currents}
\end{equation}
where the field discontinuities are defined as $\Delta\vec{E}_{t}\left(y\right)\triangleq\vec{E}_{t}^+\left(y\right)-\vec{E}_{t}^-\left(y\right)$ and $\Delta\vec{H}_{t}\left(y\right)\triangleq\vec{H}_{t}^+\left(y\right)-\vec{H}_{t}^-\left(y\right)$.

Combining Eqs. \eqref{equ:average_fields_induced_currents} and \eqref{equ:field_difference_induced_currents} yield the bianisotropic sheet transition conditions (BSTCs), relating the tangential fields below and above the metasurface to the O-BMS electric, magnetic, and magnetoelectric response. These conditions are analogous to the generalized sheet transition conditions (GSTCs) derived in \cite{Kuester2003} for anisotropic metasurfaces and utilized in \cite{Pfeiffer2013,Selvanayagam2013}, for instance, to design HMSs. For 2D configurations excited by TE-polarized fields as considered herein, the BSTCs reduce to
\begin{equation}
\left\{\!\!\!
	\begin{array}{l}
	\vspace{3pt}
	\frac{1}{2}\left(E_{x}^{+}+E_{x}^{-}\right)=-{{Z}_{se}}\left( H_{y}^{+}-H_{y}^{-} \right)-{{K}_{em}}\left( E_{x}^{+}-E_{x}^{-} \right)  \\
	\vspace{3pt}
   \frac{1}{2}\left(H_{y}^{+}+H_{y}^{-}\right)=-{{Y}_{sm}}\left( E_{x}^{+}-E_{x}^{-} \right)+{{K}_{em}}\left( H_{y}^{+}-H_{y}^{-} \right),
	\end{array}
\right.
\label{equ:BSTCs}
\end{equation}
where we have omitted the $y$ coordinate dependency for brevity. The O-BMS is passive and lossless if the impedance and admittance are purely reactive and if the magnetoelectric coupling coefficient is purely real, i.e. if $\Re\{Z_{se}\}=\Re\{Y_{sm}\}=\Im\{K_{em}\}=0$ \cite{Radi2013}.

Finally, it can be shown that if the fields below and above the metasurface satisfy local power conservation \eqref{equ:local_power_conservation} then a passive and lossless solution to the BSTCs \eqref{equ:BSTCs} can be formulated as \textcolor{black}{(Appendix \ref{app:OBMS_design_formulas})}
\begin{equation}
\left\lbrace\!\!\!
\begin{array}{l}
	\vspace{3pt}
	{{K}_{em}}=\frac{1}{2}\frac{\Re \left\{ E_{x}^{+}H{{_{y}^{-}}^{*}}-E_{x}^{-}H{{_{y}^{+}}^{*}} \right\}}{\Re \left\{ \left( E_{x}^{+}-E_{x}^{-} \right){{\left( H_{y}^{+}-H_{y}^{-} \right)}^{*}} \right\}} \\ 
	\vspace{3pt}
 {{Y}_{sm}}=-j\left( \frac{1}{2}\Im \left\{ \frac{H_{y}^{+}+H_{y}^{-}}{E_{x}^{+}-E_{x}^{-}} \right\}-{{K}_{em}}\Im \left\{ \frac{H_{y}^{+}-H_{y}^{-}}{E_{x}^{+}-E_{x}^{-}} \right\} \right) \\ 
 \vspace{3pt}
 {{Z}_{se}}=-j\left( \frac{1}{2}\Im \left\{ \frac{E_{x}^{+}+E_{x}^{-}}{H_{y}^{+}-H_{y}^{-}} \right\}+{{K}_{em}}\Im \left\{ \frac{E_{x}^{+}-E_{x}^{-}}{H_{y}^{+}-H_{y}^{-}} \right\} \right). 
\end{array}
\right.
\label{equ:passive_lossless_design}
\end{equation}
In other words, if we consider a scenario in which all the sources reside below the metasurface (Fig. \ref{fig:physical_configuration}), then the passive and lossless O-BMS given by \eqref{equ:passive_lossless_design} will convert the given source fields $\left\{E_x^<\left(y,z\right),H_y^<\left(y,z\right),H_z^<\left(y,z\right)\right\}$ to the desirable transmitted fields $\left\{E_x^>\left(y,z\right),H_y^>\left(y,z\right),H_z^>\left(y,z\right)\right\}$, provided that this field transformation satisfies \eqref{equ:local_power_conservation}.

\subsection{Microscopic (meta-atom) design}
\label{subsec:micro_design}
After the required (macroscopic) surface constituents have been evaluated via \eqref{equ:passive_lossless_design}, a suitable physical structure is to be devised for implementing the desirable O-BMS. As discussed in Section \ref{sec:introduction}, our goal is to come up with a general meta-atom configuration which exhibits omega-type bianisotropy, while being compatible with standard fabrication techniques.

To \textcolor{black}{this} end, we examine the local O-BMS properties from a microwave network perspective, and develop a suitable circuit model for such a meta-atom \cite{Pozar2012}. Accordingly, we consider an infinite periodic array of identical unit cells, corresponding to the O-BMS parameters of \eqref{equ:passive_lossless_design} at a given point $y=y_0$, namely, $Z_{se}$, $Y_{sm}$, and $K_{em}$. Based on the principle of local periodicity, we will assume that the local properties of the O-BMS at $y=y_0$ can be approximated by the scattering properties of this infinite periodic array \cite{Epstein2015_3}. 

\begin{figure}[!t]
\centering
\includegraphics[width=8cm]{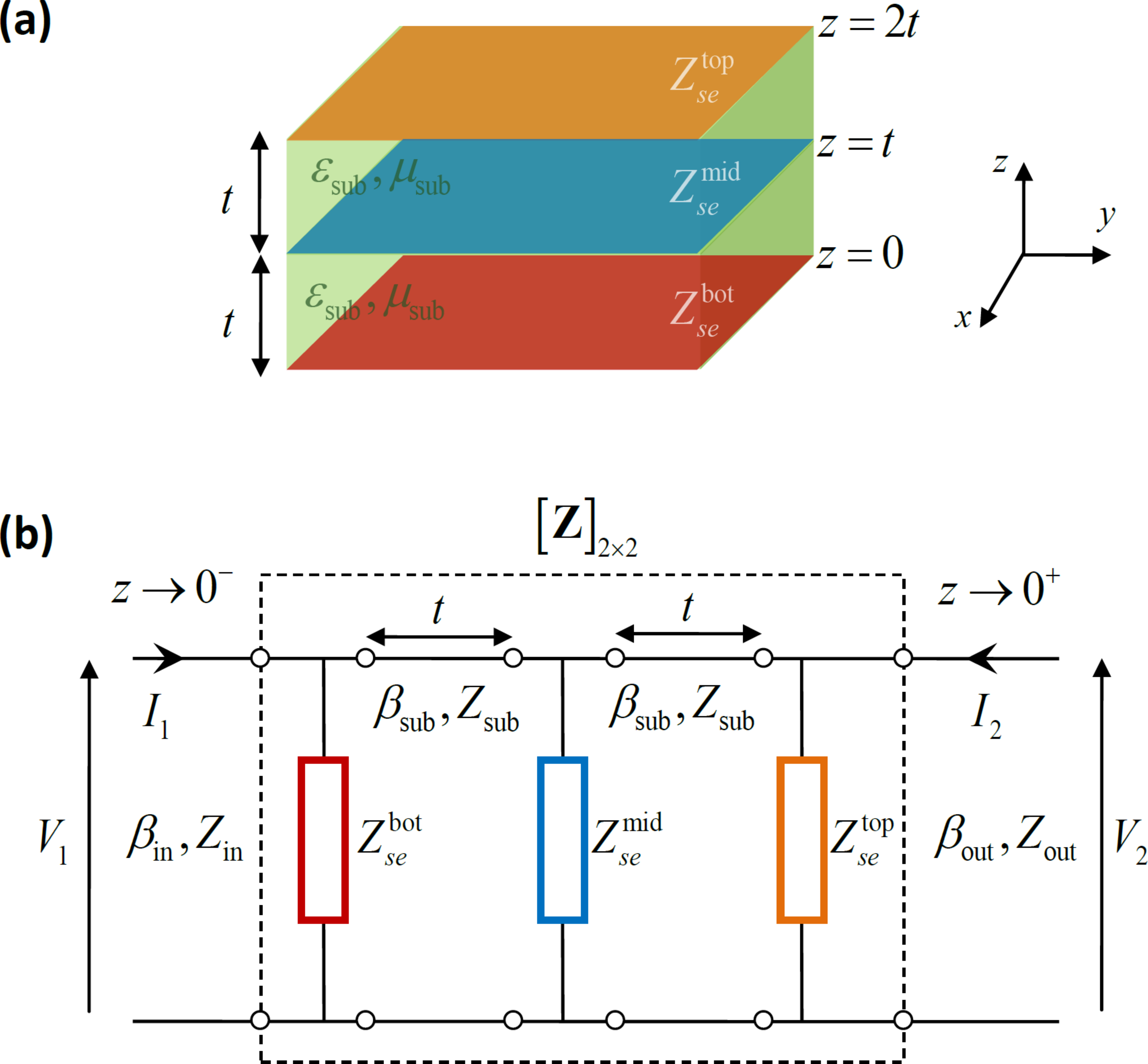}%
\caption{(a) Three-layer impedance sheet structure implementing a general omega-type bianisotropic meta-atom. (b) The corresponding equivalent circuit model. An infinite periodic array of subwavelength identical unit cells can be characterized by a $2\times 2$ impedance matrix $\mathbf{[Z]}$ relating the voltage and current at port 1 and port 2 in a transmission line model, corresponding, respectively, to the tangential electric and magnetic fields just below and above the O-BMS. This model is harnessed for the microscopic design of the O-BMS, where the meta-atom is realized by an asymmetric cascade of three impedance sheets, $Z_{se}^\mathrm{bot}$, $Z_{se}^\mathrm{mid}$, and $Z_{se}^\mathrm{top}$, implementing the required impedance matrix.}
\label{fig:circuit_model}
\end{figure}

This analogy is very useful, as it allows us to treat a scalar O-BMS unit cell as a two-port microwave network, characterized by a $2\times2$ impedance matrix $\mathbf{[Z]}$, where the currents and voltages at ports $1$ and $2$ correspond, respectively, to the local magnetic and electric fields at the bottom and top facets of the metasurface \cite{Selvanayagam2013_2,Epstein2015_3} (Fig. \ref{fig:circuit_model}). In this transmission line model\textcolor{black}{,} the electric and magnetic fields (or the equivalent voltages and currents) are related via a characteristic impedance, dependent of the nature of the excitation/scattering at $y=y_0$. 

Similar to the BSTCs \eqref{equ:BSTCs}, the impedance matrix defines relations between the tangential fields above and below the metasurface. By definition, these are given by
\begin{equation}
\left(
\begin{array}{c} \vspace{3pt}
E_x^- \\ E_x^+	
\end{array}
\right) = 
\left(
\begin{array}{c c} \vspace{3pt}
Z_{11} & Z_{12}	\\
Z_{21} & Z_{22}
\end{array}
\right)
\left(
\begin{array}{c} 
H_y^- \\ -H_y^+	
\end{array}
\right),
\label{equ:OBMS_Z_matrix}
\end{equation}
where $Z_{ij}$ are the components of the matrix $\mathbf{[Z]}$ (note that the electric surface impedance $Z_\mathrm{se}$ in \eqref{equ:BSTCs} is a scalar, unrelated to the impedance matrix $\mathbf{[Z]}$). By equating \eqref{equ:OBMS_Z_matrix} with \eqref{equ:BSTCs}, it is possible to express the impedance matrix components as a function of the O-BMS constituents (similar to \cite{Selvanayagam2013_2, Epstein2015_3}). This yields
\begin{equation}
\left\lbrace\!\!\!
\begin{array}{l}
	\vspace{3pt}
	Z_{11}=Z_{se}+\dfrac{\left(1+2K_{em}\right)^2}{4Y_{sm}} \\
	\vspace{3pt}
	Z_{12}=Z_{21}=Z_{se}-\dfrac{\left(1-2K_{em}\right)\left(1+2K_{em}\right)}{4Y_{sm}} \\
	\vspace{3pt}
	Z_{22} = Z_{se}+\dfrac{\left(1-2K_{em}\right)^2}{4Y_{sm}}.
\end{array}
\right.
\label{equ:OBMS_Z_matrix_explicit}
\end{equation}

This result is consistent with the impedance matrix corresponding to a Huygens' metasurface \cite{Selvanayagam2013_2,Epstein2015_3}, reproduced by substituting $K_{em}=0$ into \eqref{equ:OBMS_Z_matrix_explicit}, as expected (by definition, HMSs have a negligible magnetoelectric coupling). The latter finding highlights a primary difference between the equivalent circuits of the two classes of meta-atoms: while for HMSs $Z_{11}=Z_{22}$, corresponding to a symmetric structure (with respect to the $\widehat{xy}$ plane), for O-BMSs $Z_{11}\neq Z_{22}$, corresponding to an asymmetric structure \cite{Wong2015_1}. \textcolor{black}{It has} been previously shown that HMS unit cells can be implemented by cascading three \emph{symmetric} electric-impedance sheets \cite{Monticone2013,Pfeiffer2013_1,Pfeiffer2014}\textcolor{black}{; thus,} it would be natural to attempt to associate an \emph{asymmetric} cascade of three electric-impedance sheets with an O-BMS unit cell. 

We thus consider three infinite electric-impedance sheets $Z_\mathrm{se}^\mathrm{bot}$, $Z_\mathrm{se}^\mathrm{mid}$, $Z_\mathrm{se}^\mathrm{top}$ positioned at $z=0$, $z=t$, and $z=2t$, respectively, separated by a dielectric substrate with permittivity $\epsilon_\mathrm{sub}$ and permeability $\mu_\mathrm{sub}$ [Fig. \ref{fig:circuit_model}(a)]. This dielectric slab acts as a transmission line (TL) for normally-incident plane waves (transverse-electric-magnetic (TEM) modes), with a corresponding wave impedance of $Z_\mathrm{sub}=\sqrt{\mu_\mathrm{sub}/\epsilon_\mathrm{sub}}$ and a longitudinal wavenumber of $\beta_\mathrm{sub}=\omega\sqrt{\mu_\mathrm{sub}\epsilon_\mathrm{sub}}$ [Fig. \ref{fig:circuit_model}(b)]. Therefore, the scattering problem of a normally-incident plane wave impinging upon the three-layered structure can be solved using the corresponding TL model, yielding a closed-form expression for the equivalent impedance matrix, depending on the values of the impedance sheets \cite{FelsenMarcuvitz1973,Monticone2013,Wong2015_1,Pfeiffer2014_3}. By inverting this parametric expression, it is possible to derive the required (passive and lossless) $Z_\mathrm{se}^\mathrm{bot}$, $Z_\mathrm{se}^\mathrm{mid}$, $Z_\mathrm{se}^\mathrm{top}$ to implement a given $\mathbf{[Z]}$; these read \cite{Wong2015_1}
\begin{equation}
\left\lbrace\!\!\!
\begin{array}{l}
	\vspace{3pt}
	Z_\mathrm{se}^\mathrm{bot}=\dfrac{Z_\mathrm{sub}\tan\left(\beta_\mathrm{sub}t\right)}{j+Z_\mathrm{sub}\tan\left(\beta_\mathrm{sub}t\right)\frac{Z_{11}+Z_{12}}{\Delta_Z}} \\
	\vspace{3pt}	\textcolor{black}{Z_\mathrm{se}^\mathrm{mid}=-\dfrac{\left[Z_\mathrm{sub}\tan\left(\beta_\mathrm{sub}t\right)\right]^2\frac{Z_{12}}{\Delta_Z}}{\sec^2\left(\beta_\mathrm{sub}t\right)-2jZ_\mathrm{sub}\tan\left(\beta_\mathrm{sub}t\right)\frac{Z_{12}}{\Delta_Z}}} \\
	\vspace{3pt}
	Z_\mathrm{se}^\mathrm{top} = \dfrac{Z_\mathrm{sub}\tan\left(\beta_\mathrm{sub}t\right)}{j+Z_\mathrm{sub}\tan\left(\beta_\mathrm{sub}t\right)\frac{Z_{22}+Z_{12}}{\Delta_Z}},
\end{array}
\right.
\label{equ:OBMS_impedance_sheets}
\end{equation}
where $\Delta_Z=Z_{11}Z_{22}-Z_{12}^2$ is the determinant of the matrix $\mathbf{[Z]}$. It should be noted that while the BSTCs \eqref{equ:BSTCs} are defined on the plane $z=0$, the matrix $\mathbf{[Z]}$ corresponds to a finite structure, defining the transition relations between the fields at $z=0$ and $z=2t$. This means that the transmitted fields on the metasurface ($z\rightarrow 0^+$) should be properly deembedded to the plane $z=2t$ when the relations \eqref{equ:passive_lossless_design}-\eqref{equ:OBMS_impedance_sheets} are used.

It is worth pointing out that although this cascaded unit cell is prone, in general, to spatial dispersion (i.e., the impedance matrix is dependent on the angle of incidence of the exciting plane wave), this effect can be minimized by reducing the substrate thickness $t$ \cite{Epstein2015_3}. This can be deduced from \eqref{equ:OBMS_impedance_sheets}: when considering a small electrical thickness $\beta_\mathrm{sub}t\ll 1$ the expressions for the required sheet impedances are reduced to
\begin{equation}
\left\lbrace\!\!\!
\begin{array}{l}
	\vspace{3pt}
	Z_\mathrm{se}^\mathrm{bot}=\dfrac{Z_\mathrm{sub}\beta_\mathrm{sub}t}{j+Z_\mathrm{sub}\beta_\mathrm{sub}t\frac{Z_{11}+Z_{12}}{\Delta_Z}} \\
	\vspace{3pt}	\textcolor{black}{Z_\mathrm{se}^\mathrm{mid}=-\dfrac{\left(Z_\mathrm{sub}\beta_\mathrm{sub}t\right)^2\frac{Z_{12}}{\Delta_Z}}{1-2jZ_\mathrm{sub}\beta_\mathrm{sub}t\frac{Z_{12}}{\Delta_Z}}} \\
	\vspace{3pt}
	Z_\mathrm{se}^\mathrm{top} = \dfrac{Z_\mathrm{sub}\beta_\mathrm{sub}t}{j+Z_\mathrm{sub}\beta_\mathrm{sub}t\frac{Z_{22}+Z_{12}}{\Delta_Z}}.
\end{array}
\right.
\label{equ:OBMS_impedance_sheets_small_thickness}
\end{equation}
For the considered TE-polarized excitation the wave-impedance/longitudinal-wavenumber product is independent of the angle of incidence; thus, the factor $Z_\mathrm{sub}\beta_\mathrm{sub}=\omega\mu_\mathrm{sub}$ will retain its value for arbitrary plane-wave excitations. In other words, if the substrate is sufficiently thin, the impedance matrix corresponding to the cascaded structure defined by \eqref{equ:OBMS_impedance_sheets} will exhibit only minor spatial dispersion. 

We are now ready to outline the complete O-BMS design procedure. First, the desirable fields $\left\{E_x^<\left(y,z\right),H_y^<\left(y,z\right)\right\}$ and $\left\{E_x^>\left(y,z\right),H_y^>\left(y,z\right)\right\}$ (below and above the metasurface) are stipulated; these fields must be compatible with Maxwell's equations and the relevant boundary, continuity, source, and radiation conditions. Next, we evaluate the fields at $z\rightarrow0^\pm$ and verify that they indeed satisfy local power conservation \eqref{equ:local_power_conservation}. Subsequently, we substitute the stipulated fields into \eqref{equ:passive_lossless_design}, yielding the required (macroscopic) O-BMS specifications, corresponding to a passive and lossless design. We then sample the resulting $K_{em}\left(y\right)$, $X_{se}\left(y\right)\triangleq\Im\left\{Z_{se}\left(y\right)\right\}$, and $B_{sm}\left(y\right)\triangleq\Im\left\{Y_{sm}\left(y\right)\right\}$ at discrete points according to the meta-atom lateral dimensions, and evaluate the local impedance matrix $\mathbf{[Z]}\left(y\right)$ at each point following \eqref{equ:OBMS_Z_matrix_explicit}. Finally, we implement each such unit cell as an asymmetric cascade of three impedance sheets, following \eqref{equ:OBMS_impedance_sheets} with the required $\mathbf{[Z]}\left(y\right)$ and the suitable substrate parameters $Z_\mathrm{sub}$, $\beta_\mathrm{sub}$, and $t$.

In the subsequent section we will demonstrate the utilization of this procedure for the design of O-BMSs for various applications. The full-wave simulation results presented therein are obtained by defining meta-atoms as in Fig. \ref{fig:circuit_model}(a) in ANSYS HFSS, using the impedance boundary condition feature to realize the impedance sheets prescribed by \eqref{equ:OBMS_impedance_sheets}. We use two Rogers RO3010 laminates of $t=5\mathrm{mil}$ thickness as \textcolor{black}{the} substrate in our meta-atom. These commercially available products feature permittivity of $\epsilon_\mathrm{sub}=13.06\epsilon_0$ at the design frequency $f=20\mathrm{GHz}$ ($\lambda\approx1.5\mathrm{cm}$), where $\epsilon_0$ is the vacuum permittivity, and a loss tangent of $\tan\delta=0.002$, incorporated into the simulation. For the evaluation of the required impedance sheets \eqref{equ:OBMS_impedance_sheets} we neglect these small dielectric losses, thus using $Z_\mathrm{sub}=0.2767\eta$, $\beta_\mathrm{sub}=3.613k$, and $t=\lambda/118$, where the O-BMSs considered to be embedded in vacuum (Fig. \ref{fig:physical_configuration}). For this choice of parameters, the electrical length of the spacers is $\beta_\mathrm{sub}t=0.19$, which is sufficient to ensure minimal spatial dispersion of the meta-atoms \eqref{equ:OBMS_impedance_sheets_small_thickness}. \textcolor{black}{The surface properties of \eqref{equ:passive_lossless_design} are discretized along the $y$ dimension and implemented by unit cells of length $\Delta_y=\lambda/9.5\approx1.58\mathrm{mm}$, a realistic size for microwave metasurfaces \cite{Wong2014, Epstein2016}. Nonetheless, depending on the application and the expected variation of the O-BMS constituents, using larger unit cell sizes may be possible; for the case considered in Subsection \ref{subsec:refraction}, for instance, we have noticed that a discretization of $\lambda/5$ does not significantly affect the device performance.}

\textcolor{black}{Several comments are in place with regards to the presented unit-cell design scheme. First, although we consider herein a three-layer structure for implementing the O-BMS meta-atoms, there could be other options to realize them. The O-BSTCs imply that the unit cells should feature \emph{three} degrees of freedom, allowing simultaneous realization of the desirable electric surface impedance, magnetic surface admittance, and magnetoelectric coupling coefficient values. Nonetheless, our proposed configuration and circuit model (Fig. \ref{fig:circuit_model}) actually includes \emph{six} degrees of freedom: three sheet reactances, the substrate thickness $t$, and the substrate constituents, $\epsilon_\mathrm{sub}$ and $\mu_\mathrm{sub}$. Therefore, it is quite plausible that the required omega-bianisotropy could be achieved by using asymmetric formations with only \emph{two} metallic layers (i.e. two reactive sheets), and proper tuning of the substrate thickness and properties, as implied by \cite{Yazdi2015,Alaee2015}, for instance. However, when considering a general inhomogeneous metasurface (\textit{cf.} Subsections \ref{subsec:refraction} and \ref{subsec:cavity_excited_antenna}) and standard PCB-compatible fabrication processes, locally-modifying the geometry of metallic layers seems more practical than local variation of the substrate thickness or electromagnetic constituents. Thus, although for specific applications simpler structures may meet the design requirements, we have adopted the presented three-layer meta-atom configuration to be able to accommodate the most general scenarios. 
}

\textcolor{black}{Second, while the circuit model of Fig. \ref{fig:circuit_model} and the corresponding formulation indicate that three cascaded impedance sheets feature the physical mechanisms required for implementing omega-type meta-atoms, and also allow numerical verification via commercially-available solvers (a non-trivial task by itself), additional steps are necessary to achieve fabrication-ready designs. In practical designs, the abstract sheet impedances are replaced with suitable copper traces, and inter-coupling between adjacent layers, which is not accounted for in the model, may become significant \cite{Pfeiffer2013_1,Pfeiffer2013_2,Pfeiffer2014_2,Pfeiffer2014_3}. In addition, realistic structures typically include bonding layers and feature unavoidable losses \cite{Abadi2014, Epstein2016}, which further complicate the derivation. Therefore, the sheet reactance values prescribed by the model are to be used as starting points for final optimization of the conductor geometries \cite{Pfeiffer2014_3}, ensuring (via full-wave simulations) that the coupled structure complies with the overall required bianisotropic response \eqref{equ:OBMS_Z_matrix_explicit} \cite{Epstein2015_3}. Although such a detailed design is beyond the scope of this paper, we provide in Appendix \ref{app:meta_atom_physical_structure} representative examples for three-layer omega-type meta-atoms, to demonstrate the viability of this concept.   
}


\section{Results and discussion}
\label{sec:results}

\subsection{Reflectionless wide-angle plane-wave refraction}
\label{subsec:refraction}
\begin{figure}[!t]
 \centering
\includegraphics[width=8.5cm]{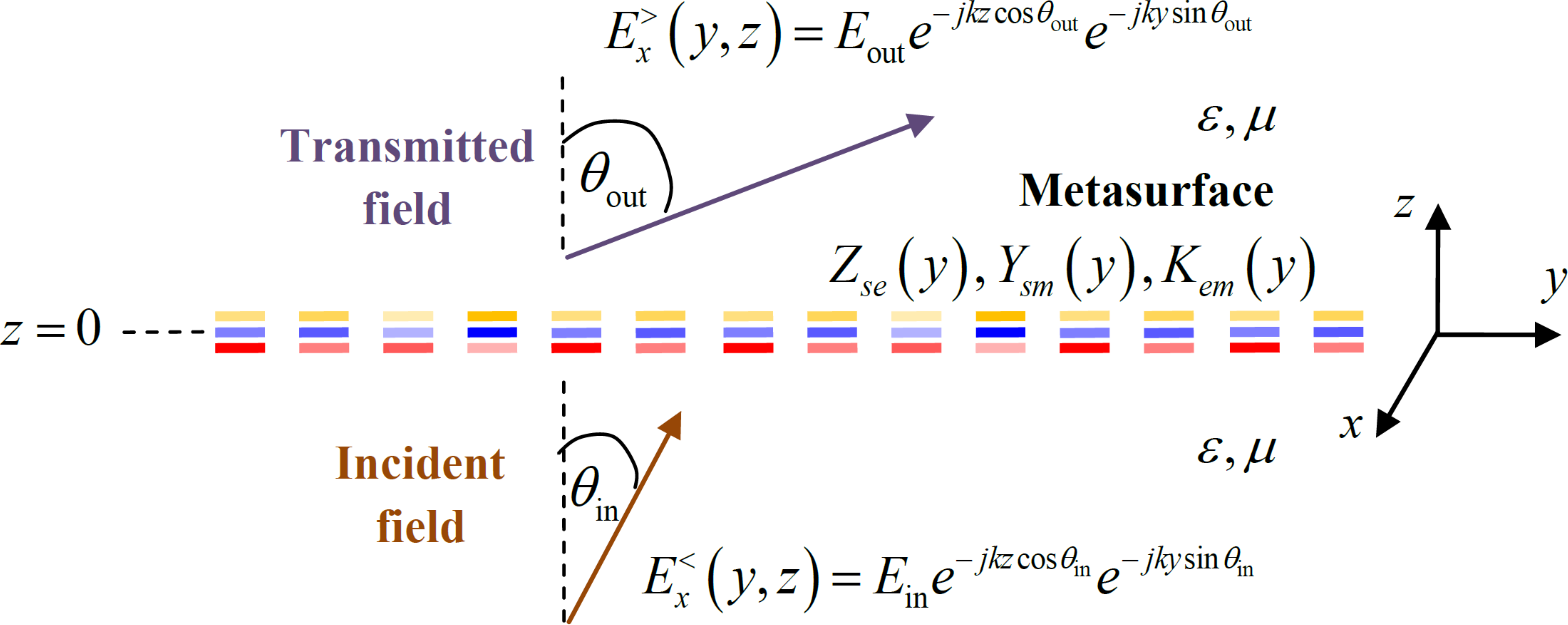}%
\caption{Physical configuration of an O-BMS implementing plane-wave refraction. All the power incident upon the O-BMS at $\theta_\mathrm{in}$ is coupled to the transmitted plane wave propagating towards $\theta_\mathrm{out}$ without incurring any reflection.}
\label{fig:physical_configuration_refraction}
\end{figure}

We begin by utilizing our derivation (Subsection \ref{subsec:macro_design}) to design an O-BMS to refract a plane wave incident at an angle of $\theta_\mathrm{in}$ towards an angle of $\theta_\mathrm{out}$ without incurring \emph{any} reflection (Fig. \ref{fig:physical_configuration_refraction}). Although for moderate angles HMSs can implement plane-wave refraction with small reflections \cite{Pfeiffer2013}, they are not capable of completely eliminating them \cite{Selvanayagam2013,Epstein2014}, and they become more significant as the difference between the incident and transmitted wave impedances becomes larger \cite{Epstein2014_2}.

In contrast, as the corresponding field transformation locally-conserves the power as required by \eqref{equ:local_power_conservation}, O-BMSs can implement a \emph{truly-reflectionless} plane-wave refraction for any given angles of incidence and transmission. Hence, to design an O-BMS that implements such a functionality we merely need to stipulate the suitable fields above and below the metasurface, and then use \eqref{equ:passive_lossless_design} to evaluate the required surface specifications. Subsequently, we define
\begin{align}
&\left\lbrace\!\!\!
\begin{array}{l}
	\vspace{3pt}
	E_x^<\left(y,z\right) = E_\mathrm{in}e^{-jk\cos\theta_\mathrm{in}z}e^{-jk\sin\theta_\mathrm{in}y} \\
	\vspace{3pt}
	H_y^<\left(y,z\right) = \frac{1}{Z_\mathrm{in}}E_\mathrm{in}e^{-jk\cos\theta_\mathrm{in}z}e^{-jk\sin\theta_\mathrm{in}y} 
\end{array}
\right. \nonumber \\
&\left\lbrace\!\!\!
\begin{array}{l}
	\vspace{3pt}
	E_x^>\left(y,z\right) = E_\mathrm{out}e^{-jk\cos\theta_\mathrm{out}z}e^{-jk\sin\theta_\mathrm{out}y} \\
	\vspace{3pt}
	H_y^>\left(y,z\right) =\frac{1}{Z_\mathrm{out}} E_\mathrm{out}e^{-jk\cos\theta_\mathrm{out}z}e^{-jk\sin\theta_\mathrm{out}y},
\end{array}
\right.
\label{equ:fields_plane_wave_refraction}
\end{align}
where $Z_\mathrm{in}=1/Y_\mathrm{in}\triangleq\eta/\cos\theta_\mathrm{in}$ and $Z_\mathrm{out}=1/Y_\mathrm{out}\triangleq\eta/\cos\theta_\mathrm{out}$ are the wave impedances of TE-polarized plane waves propagating at $\theta_\mathrm{in}$ and $\theta_\mathrm{out}$ with respect to the $z$ axis \cite{FelsenMarcuvitz1973}; $E_\mathrm{in}$ and $E_\mathrm{out}$ are the amplitudes of the incident and transmitted plane waves, respectively.

In order to utilize the methodology developed in Subsection \ref{subsec:macro_design}, we must first ensure that local power conservation \eqref{equ:local_power_conservation} is satisfied. This is achieved by setting the amplitude of the transmitted wave following
\begin{equation}
E_\mathrm{out}=\sqrt{\frac{Z_\mathrm{out}}{Z_\mathrm{in}}}\left|E_\mathrm{in}\right|e^{-j\xi_\mathrm{out}},
\label{equ:power_conservation_refraction}
\end{equation}
where $\xi_\mathrm{out}$ is an arbitrary uniform phase-shift that can be added to the transmitted plane wave, if desirable. Once this condition is met, all is left to do is substitute \eqref{equ:fields_plane_wave_refraction} into \eqref{equ:passive_lossless_design}, which yields the required O-BMS properties. These are given by 
\begin{equation}
\left\lbrace\!\!\!
\begin{array}{l}
	\vspace{3pt}
	{{K}_{em}}\left(y\right)=\dfrac{\Delta Z}{4\overline{Z}_\mathrm{G}}\dfrac{\cos \left( ky{{\Delta }_{\sin }}+{{\xi }_\mathrm{out}} \right)}{1-\left( \overline{Z}_\mathrm{A}/\overline{Z}_\mathrm{G} \right)\cos \left( ky{{\Delta }_{\sin }}+{{\xi }_\mathrm{out}} \right)} \\ 
	\vspace{3pt}
 {Y}_{sm}\left(y\right)=-j\dfrac{\overline{Y}_\mathrm{G}}{2}\dfrac{\sin \left( ky{{\Delta }_{\sin }}+{{\xi }_\mathrm{out}} \right)}{1-\left( \overline{Z}_\mathrm{A}/\overline{Z}_\mathrm{G}\right)\cos \left( ky{{\Delta }_{\sin }}+{{\xi }_\mathrm{out}} \right)}  \\ 
 \vspace{3pt}
 {Z}_{se}\left(y\right)=-j\dfrac{\overline{Z}_\mathrm{G}}{2}\dfrac{\sin \left( ky{{\Delta }_{\sin }}+{{\xi }_\mathrm{out}} \right)}{1-\left( \overline{Z}_\mathrm{A}/\overline{Z}_\mathrm{G}\right)\cos \left( ky{{\Delta }_{\sin }}+{{\xi }_\mathrm{out}} \right)}, 
\end{array}
\right.
\label{equ:OBMS_specifications_refraction}
\end{equation}
where $\overline{Z}_\mathrm{A}\triangleq\left(Z_\mathrm{out}+Z_\mathrm{in}\right)/2$, $\overline{Z}_\mathrm{G}=1/\overline{Y}_\mathrm{G}\triangleq\sqrt{Z_\mathrm{out}Z_\mathrm{in}}$, and $\Delta Z\triangleq Z_\mathrm{out}-Z_\mathrm{in}$ are, respectively, the arithmetic mean, geometric mean, and difference, of the output and input wave impedances; $\Delta_{\sin}\triangleq\sin\theta_\mathrm{out}-\sin\theta_\mathrm{in}$.

Equation \eqref{equ:OBMS_specifications_refraction} reveals the role of the magnetoelectric coupling in establishing a reflectionless refracting metasurface. The corresponding coefficient $K_{em}$ is seen to be proportional to the impedance mismatch $\Delta Z/\overline{Z}_\mathrm{G}$, indicating that the omega-type bianisotropy facilitates the matching between the incident and transmitted wave impedances. When this mismatch tends to zero, i.e when $\theta_\mathrm{out}\rightarrow\pm\theta_\mathrm{in}$, the required magnetoelectric coupling becomes negligible, and the metasurface constituents approach those of a refracting Huygens' metasurface \cite{Selvanayagam2013, Epstein2014}. Indeed, when $\Delta Z=0$, HMSs can also implement reflectionless refraction; however, when the mismatch is significant, passive lossless HMSs would usually require reflected fields to ensure local wave-impedance equalization \cite{Epstein2014, Epstein2014_2}.

This point is further emphasized when the impedance matrix corresponding to the O-BMS is examined. For the specific case of plane-wave refraction, $\mathbf{[Z]}$ is evaluated by substituting the O-BMS parameters derived in \eqref{equ:OBMS_specifications_refraction} into \eqref{equ:OBMS_Z_matrix_explicit}, yielding
\begin{equation}
\left\lbrace\!\!\!
\begin{array}{l}
	\vspace{3pt}
	Z_{11}=-jZ_\mathrm{in}\cot\left(ky\Delta_{\sin}+\xi_\mathrm{out}\right) \\
	\vspace{3pt}
	Z_{12}=Z_{21}=-j\dfrac{\overline{Z}_\mathrm{G}}{\sin\left(ky\Delta_{\sin}+\xi_\mathrm{out}\right)} \\
	\vspace{3pt}
	Z_{22} = -jZ_\mathrm{out}\cot\left(ky\Delta_{\sin}+\xi_\mathrm{out}\right).
\end{array}
\right.
\label{equ:Z_matrix_refraction}
\end{equation}
To clarify the physical meaning of such an impedance matrix, we transform it into a \emph{generalized} scattering matrix $\mathbf{[G]}$ with input and output ports having characteristic impedances $Z_\mathrm{in}$ and $Z_\mathrm{out}$, respectively \cite{Frickey1994}. Such a matrix describes how field wavefronts with \emph{different} wave impedances scatter off a given system \cite{Pozar2012}, in complete analogy to the scenario under consideration. Subsequently, the components $G_{11}$ and $G_{22}$, respectively, correspond to the reflection coefficients for plane waves with wave impedances $Z_\mathrm{in}$ and $Z_\mathrm{out}$ propagating below or above the metasurface (Fig. \ref{fig:circuit_model}); the components $G_{12}$ and $G_{21}$, which are identical in our reciprocal configuration, correspond to the coefficient imposed on the incident wave fields upon transmission through the metasurface. 

Executing the algebraic transformation from $\mathbf{[Z]}$ to $\mathbf{[G]}$ we arrive at \cite{Frickey1994}
\begin{equation}
\left\lbrace\!\!\!
\begin{array}{l}
	\vspace{3pt}
	G_{11}=G_{22}=0 \\
	\vspace{3pt}
	G_{12}=G_{21}=e^{-jky\Delta_\mathrm{sin}}e^{-j\xi_\mathrm{out}}.
\end{array}
\right.
\label{equ:G_matrix_refraction}
\end{equation}
This verifies that the O-BMS design presented in \eqref{equ:OBMS_specifications_refraction} reflects different wave impedances to its input ($z\rightarrow0^-$) and output ($z\rightarrow0^+$) ports, allowing for perfect matching (zero reflection) of the metasurface to plane waves propagating in independent directions below and above it. This capability is directly related to the asymmetry introduced by the magnetoelectric coupling, manifested in the impedance matrix by the different values of $Z_{11}$ and $Z_{22}$.

It is worth noting that the formulation developed herein provides rigorous justification to the reflectionless refracting metasurface introduced recently in \cite{Wong2015_1}. In that work, Wong \textit{et al.} hypothesize that in order to implement truly-reflectionless wide-angle refraction, each of the metasurface unit cells has to be impedance matched to both the incident and transmitted waves. Based on that hypothesis they devised a design based on the generalized scattering matrix $\mathbf{[G]}$ of \eqref{equ:G_matrix_refraction}, transforming it to $\mathbf{[Z]}$ of \eqref{equ:Z_matrix_refraction}. The required impedance matrix was shown to correspond to an asymmetric stack of three impedance sheets (as in Fig. \ref{fig:circuit_model}), and an equivalent realization of the metasurface in ANSYS HFSS was carried. Subsequently, it was verified through full-wave simulations that, indeed, wide-angle refraction can be achieved without any reflection whatsoever (refraction from $\theta_\mathrm{in}=0$ to $\theta_\mathrm{out}=80^\circ$ was demonstrated with $98.89\%$ of the power coupled to the refracted wave \cite{Wong2015_1}). The relation between the impedance matrix and the BSTCs presented in this subsection reveals that the structure designed therein is actually an O-BMS, and proves, by solving Maxwell's equations with the corresponding boundary conditions, that it indeed implements reflectionless (arbitrary) plane-wave refraction. 


To verify the theory, we have followed the macroscopic design procedure outlined in \eqref{equ:fields_plane_wave_refraction}-\eqref{equ:OBMS_specifications_refraction} to derive the O-BMS specifications to implement refraction of a plane wave incident at $\theta_\mathrm{in}=0$ towards $\theta_\mathrm{out}=71.81^\circ$, with an additional uniform phase shift of $\xi_\mathrm{out}=70^\circ$. We have then utilized the microscopic design procedure of Subsection \ref{subsec:micro_design} to transform these specifications to three-layer sheet-impedance meta-atoms \eqref{equ:OBMS_impedance_sheets} based on the $t=5\mathrm{mil}$ Rogers RO3010 substrates, and defined the resultant O-BMS in ANSYS HFSS accordingly. As for the chosen refraction parameters the O-BMS surface properties of \eqref{equ:OBMS_specifications_refraction} are periodic with a period of $\lambda/\left|\Delta_{\sin}\right|=\lambda/0.95$, we used Floquet ports with periodic boundary conditions to simulate the plane-wave scattering off a $10$ unit cell formation (recall that the unit cell size along the $y$ dimension is $\Delta_y=\lambda/9.5$).

 \begin{figure}[!t]
 \centering
\includegraphics[width=6cm]{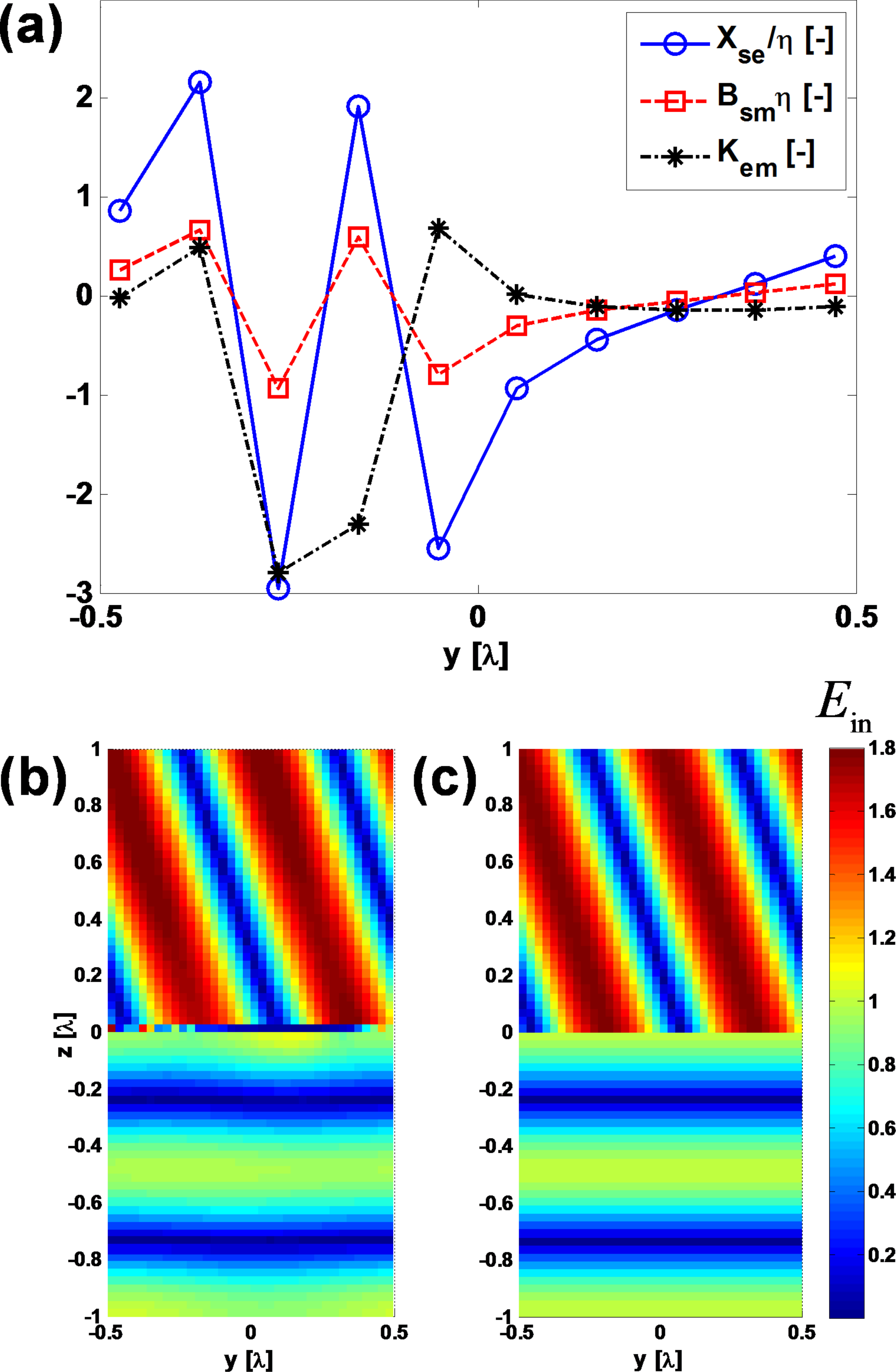}%
\caption{Plane-wave scattering off an O-BMS designed to implement refraction of a normally-incident plane wave towards $\theta_\mathrm{out}=71.81^\circ$, inducing an additional uniform phase shift of $\xi_\mathrm{out}=70^\circ$. (a) Required electric surface reactance $X_{se}$ (blue solid lines and circles), magnetic surface susceptance $B_{sm}$ (red dashed lines and squares), and magnetoelectric coupling coefficient $K_{em}$ (black dash-dotted lines and asterisks), as specified by \eqref{equ:OBMS_specifications_refraction}. (b) Simulated electric field distribution $\left|\Re\left\{E_x\left(y,z\right)\right\}\right|$. (c) Analytical prediction of $\left|\Re\left\{E_x\left(y,z\right)\right\}\right|$ via \eqref{equ:fields_plane_wave_refraction}.}
\label{fig:results_refraction}
\end{figure}

Figure \ref{fig:results_refraction} presents the specifications of the refracting O-BMS\footnote{\textcolor{black}{As the O-BMS design is fully-characterized by $X_{se}$, $B_{sm}$, and $K_{em}$, regardless of the unit cell implementation, we present these parameters for the various devices presented herein. The specific values of $Z_{se}^\mathrm{bot}$, $Z_{se}^\mathrm{mid}$, and $Z_{se}^\mathrm{top}$, on the other hand, are less general, as they depend on the particular substrate thickness $t$ and permeability $\epsilon_\mathrm{sub}$ chosen for the implementation. Appendix \ref{app:meta_atom_physical_structure} demonstrates the realizability of the metasurface specified in Fig. \ref{fig:results_refraction}(a) via three-layer impedance sheet meta-atoms by characterizing possible physical structures for representative unit cells.}} and the results of normally-incident plane-wave scattering off the metasurface. As can be seen from the electric field snapshots, the field distribution obtained from full-wave simulations [Fig. \ref{fig:results_refraction}(b)] is in excellent quantitative agreement with the one predicted from the analytical expressions of \eqref{equ:fields_plane_wave_refraction} [Fig. \ref{fig:results_refraction}(c)]. The simulated Floquet analysis reveals that $99.5\%$ of the incident power is coupled to the first Floquet-Bloch mode in transmission (propagating towards $\theta_\mathrm{out}=71.81^\circ$) verifying that, indeed, the designed O-BMS is capable of implementing truly-reflectionless wide-angle refraction. The simulation indicates that the phase-shift introduced by the metasurface corresponds to $\xi_\mathrm{out}=63.4^\circ$, a deviation of less than $2\%$ than the designated value, with respect to a full cycle. For comparison, ideal HMSs designed following \cite{Selvanayagam2013,Epstein2014} would couple less than $73\%$ of the incident power to the desirable Floquet-Bloch mode. HMS design following the optimized scheme in \cite{Epstein2014_2} would yield $92\%$ refraction efficiency in the ideal case, however the scattered fields will contribute to additional reactive power near the metasurface due to coupling to evanescent Floquet-Bloch modes. The O-BMS performance is superior to both, indicating its potential for extreme field manipulations.


\subsection{Independent surface-wave guiding}
\label{subsec:surface_wave}
\begin{figure}[!b]
 \centering
\includegraphics[width=8.5cm]{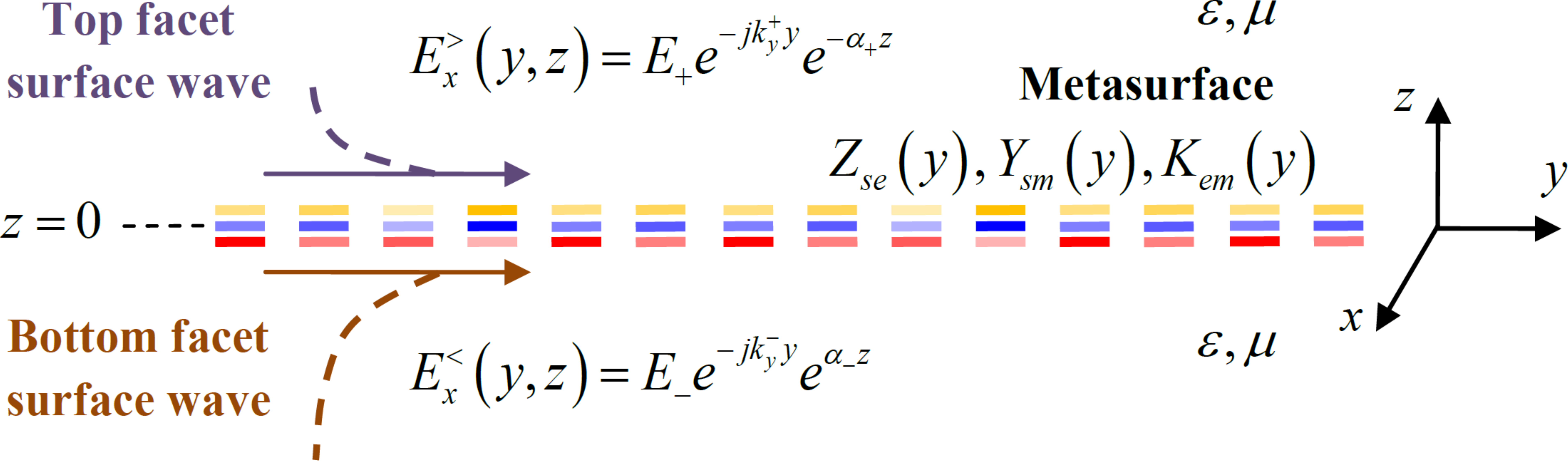}%
\caption{Physical configuration of an O-BMS supporting propagation of different surface waves on each of its facets. The surface wave guided on the top and bottom facets have decay constants of $\alpha_{+}$ and $\alpha_{-}$, respectively.}
\label{fig:physical_configuration_surface_waves}
\end{figure}

To further demonstrate the versatility of our derivation, we aim at designing a metasurface that can guide different surface waves with arbitrary propagation constants on its two facets (Fig. \ref{fig:physical_configuration_surface_waves}). Such a metasurface can be utilized to independently control the effective range of the evanescent fields extending from the metasurface into the upper and lower half-spaces, thus may be useful for applications involving near-field interactions, e.g., nonradiative wireless power transfer (NRWPT) or radio-frequency identification (RFID) \cite{Eom2009, Medeiros2011, Gonzalez2013, Morgado2014}.

\begin{figure*}[!t]
\centering
\includegraphics[width=17cm]{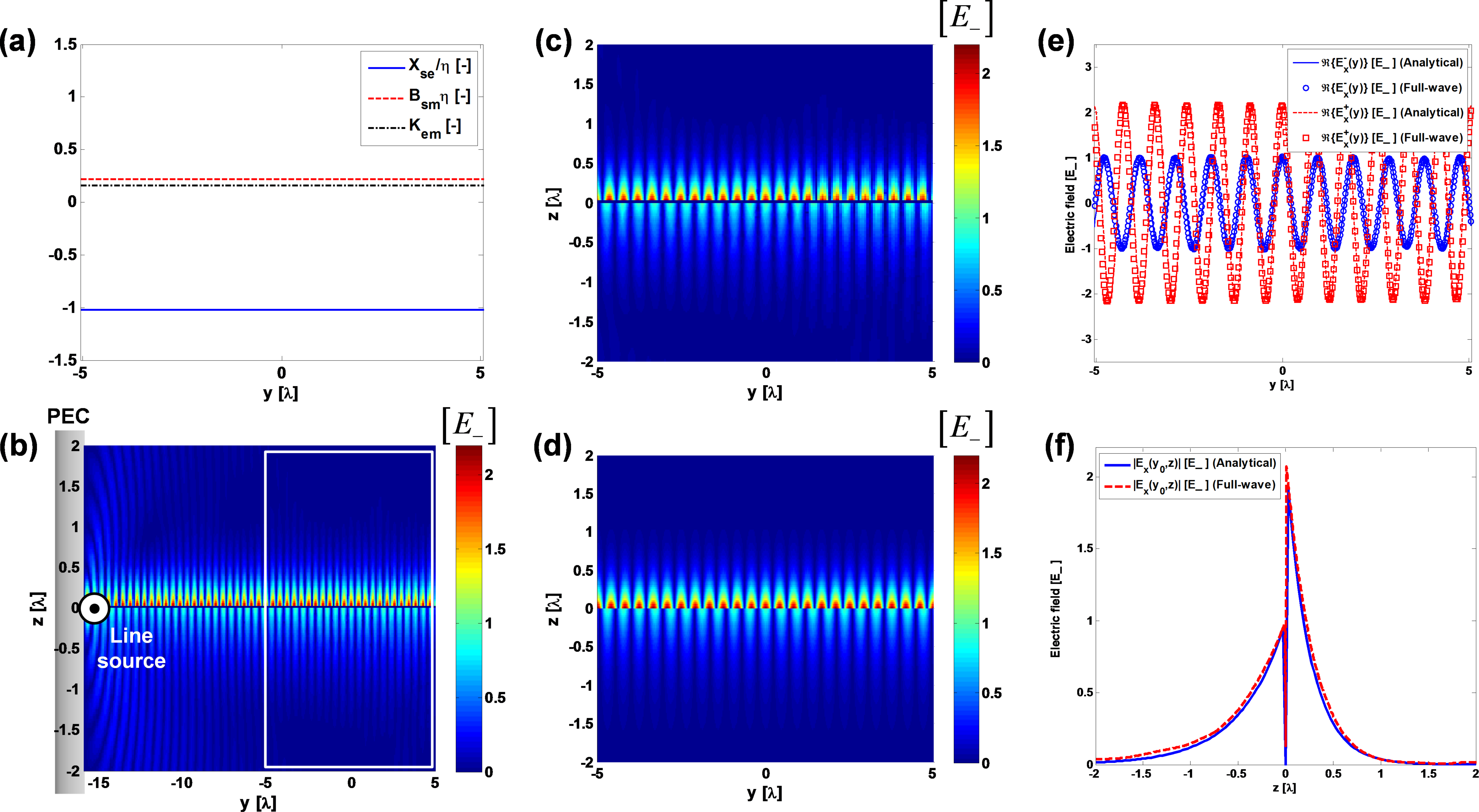}%
\caption{Excitation and propagation of independent surface waves on an O-BMS supporting surface modes with decay constants $\alpha_{-}=2.12/\lambda$ and $\alpha_{+}=4.02/\lambda$, respectively, on its bottom and top facets. (a) Required electric surface reactance $X_{se}$ (blue solid lines), magnetic surface susceptance $B_{sm}$ (red dashed lines), and magnetoelectric coupling coefficient $K_{em}$ (black dash-dotted lines), as specified by \eqref{equ:OBMS_specifications_surface_waves}. (b) Simulated electric field distribution $\left|\Re\left\{E_x\left(y,z\right)\right\}\right|$, with indication of the simulation configuration. The surface waves are excited by an electric line source positioned at $\left(y',z'\right)=\left(-16\lambda,0\right)$, in close proximity to a PEC; the planes $z=\pm 11\lambda$ implement radiation boundary conditions. The fields in the region surrounded by a white frame ($\left|y\right|<5\lambda$) are dominated by the surface wave contribution. (c) Simulated values of $\left|\Re\left\{E_x\left(y,z\right)\right\}\right|$ in the region marked in (b). (d) Analytical prediction of $\left|\Re\left\{E_x\left(y,z\right)\right\}\right|$ via \eqref{equ:fields_surface_waves} in the same region. (e) Real part of the electric field $\Re\left\{E_x\left(y,z\right)\right\}$ as a function of $y$ at $z\rightarrow0^{-}$ (blue) and $z\rightarrow0^{+}$ (red), comparing the analytical predictions via \eqref{equ:fields_surface_waves} (solid and dashed lines, respectively) to the simulated results (circles and squares, respectively). (f) Analytical prediction (solid blue line) and simulated results (dashed red line) for the electric field magnitude $\left|E_x\left(y,z\right)\right|$ as a function of $z$ at $y=y_0=5\lambda$. }
\label{fig:results_surface_waves}
\end{figure*}

As in this configuration no real power is crossing the metasurface, local power conservation \eqref{equ:local_power_conservation} is inherently satisfied, ensuring the validity of our methodology. Again, once the appropriate fields are stipulated below and above the metasurface, the O-BMS design becomes straightforward following \eqref{equ:passive_lossless_design}. For our application, we consider two surface waves propagating on the bottom and top surface of the O-BMS with different evanescent tails (in the $z$ direction), defined by $\alpha_-$ and $\alpha_+$, respectively. The tangential fields associated with such surface waves are given by
\begin{align}
&\left\lbrace\!\!\!
\begin{array}{l}
	\vspace{3pt}
	E_x^<\left(y,z\right) = E_-e^{-jk_y^-y}e^{\alpha_-z} \\
	\vspace{3pt}
	H_y^<\left(y,z\right) = j\frac{1}{\eta}\frac{\alpha_-}{k}E_-e^{-jk_y^-y}e^{\alpha_-z} 
\end{array}
\right. \nonumber \\
&\left\lbrace\!\!\!
\begin{array}{l}
	\vspace{3pt}
	E_x^>\left(y,z\right) = E_+e^{-jk_y^+y}e^{-\alpha_+z} \\
	\vspace{3pt}
	H_y^>\left(y,z\right) = -j\frac{1}{\eta}\frac{\alpha_+}{k}E_+e^{-jk_y^+y}e^{-\alpha_+z},
\end{array}
\right.
\label{equ:fields_surface_waves}
\end{align}
where $E_\pm$ are the amplitudes of the surface waves (determined by the excitation at $y=y'$) and $k_y^\pm=\sqrt{k^2+\alpha_\pm^2}$ are their propagation constants. It is worth noting that in contrast to the fields stipulated in Subsection \ref{subsec:refraction}, the local power conservation condition does not impose any constraints on the field amplitudes, and they can be independently controlled by proper engineering of the surface wave excitation sources.

To evaluate the required O-BMS constituents we substitute these fields into \eqref{equ:passive_lossless_design}, leading to the following homogeneous ($y$-invariant) specifications
\begin{equation}
\left\lbrace\!\!\!
\begin{array}{l}
	\vspace{3pt}
	{K}_{em}\left(y\right) \equiv \dfrac{1}{2}\dfrac{\alpha_+ - \alpha_-}{\alpha_+ + \alpha_-}\\ 
	\vspace{3pt}
 {Y}_{sm}\left(y\right) \equiv  \left(-j\eta\right)^{-1}\dfrac{k^{-1}}{\alpha_+^{-1}+\alpha_-^{-1}}\\
 \vspace{3pt}
 {Z}_{se}\left(y\right) \equiv -j\eta\dfrac{k}{\alpha_++\alpha_-}, 
\end{array}
\right.
\label{equ:OBMS_specifications_surface_waves}
\end{equation}
which, as expected, represent a passive and lossless O-BMS.

We utilize this methodology to design an O-BMS supporting propagation of surface waves with decay constants $\alpha_{-}=2.12/\lambda$ and $\alpha_{+}=4.02/\lambda$ on its bottom and top facets, respectively; the respective propagation constants are thus $k_y^{-}=1.056k$ and $k_y^{+}=1.188k$. Figure \ref{fig:results_surface_waves}(a) presents the corresponding O-BMS constituents \eqref{equ:OBMS_specifications_surface_waves}, which were used to implement the O-BMS in ANSYS HFSS following the microscopic design procedure of Subsection \ref{subsec:micro_design} ($t=5\mathrm{mil}$). To excite the metasurface in these full-wave simulations we used an electric line source positioned close to a PEC terminating the left edge of the O-BMS ($y=y'=-16\lambda$), as depicted in Fig. \ref{fig:results_surface_waves}(b). The line source is positioned symmetrically around the metasurface plane $z=0$, facilitating partial coupling of the source power to the two allowed surface modes. 
\textcolor{black}{From the simulation results, we found that in such a configuration, the surface waves were excited with uneven amplitudes following the relation $\left|E_{+}/E_{-}\right|\approx 2.13$.}
To enable comparison between the simulated results and the analytical predictions \eqref{equ:fields_surface_waves} we thus normalize the fields with respect to $\left|E_{-}\right|$ and use the above ratio. As the observation points move away from the source, the field due to the radiated space wave becomes negligible in the vicinity of the metasurface, and the surface wave properties can be properly examined. 

Following this reasoning, Fig. \ref{fig:results_surface_waves}(c) presents the electric field distribution at the region surrounded by a white box in Fig. \ref{fig:results_surface_waves}(b), as obtained from full-wave simulations. The corresponding analytical prediction (considering the suitable amplitude ratio) is presented in Fig. \ref{fig:results_surface_waves}(d). Besides residual interference effects related to the space wave at some regions away from the metasurface $\left|z\right|>\lambda$, the simulated and formulated field distributions show very good agreement. This conclusion is further supported by examining the electric field variation along the $y$ axis at $z\rightarrow 0^{\pm}$ [Fig. \ref{fig:results_surface_waves}(e)], and its decay profile along the $z$ axis at $y=y_0=5\lambda$ [Fig. \ref{fig:results_surface_waves}(f)]. In both plots, the data obtained from full-wave simulations follow closely the curves evaluated using the analytical expressions of \eqref{equ:fields_surface_waves}. Quantitatively, from the average peak separation in Fig. \ref{fig:results_surface_waves}(e) we evaluate the propagation constants of the simulated fields as $k_y^{-}=1.048k$ and $k_y^{+}=1.176k$, less than $1\%$ deviation from the prescribed value. These values correspond to decay factors of $\alpha_{-}=1.97/\lambda$ and $\alpha_{+}=3.89/\lambda$, about $7\%$ and $3\%$ less than the desirable parameters [similar values are extracted by fitting the decay profiles of Fig. \ref{fig:results_surface_waves}(f)]. Nonetheless, these deviations are hardly observable in Fig. \ref{fig:results_surface_waves}(f), indicating that the O-BMS achieves good control of the evanescent fields, as required.  

We further explore the properties of O-BMSs designed for surface wave guiding as per \eqref{equ:OBMS_specifications_surface_waves} by examining their impedance and scattering matrices. Plugging these O-BMS constituents into \eqref{equ:OBMS_Z_matrix_explicit} yields the corresponding $\mathbf{[Z]}$, namely,
\begin{equation}
\begin{array}{l l l}
	Z_{11}\equiv -j\eta\dfrac{k}{\alpha_{-}}, &
	Z_{12}=Z_{21}\equiv0, &
	Z_{22} \equiv -j\eta\dfrac{k}{\alpha_{+}}.
\end{array}
\label{equ:Z_matrix_surface_waves}
\end{equation}
As in Subsection \ref{subsec:refraction}, we use this impedance matrix to construct the associated generalized scattering parameters, for plane waves with some given wave impedances $Z_\mathrm{in}$ and $Z_\mathrm{out}$, respectively, below and above the metasurface. Interestingly, these are found to be
\begin{equation}
\begin{array}{l l l}
	G_{11}\equiv e^{2j\psi_{-}}, &
	G_{12}=G_{21}\equiv0, &
	G_{22} \equiv e^{2j\psi_{+}},
\end{array}
\label{equ:G_matrix_surface_waves}
\end{equation}
where the phase shifts of $G_{11}$ and $G_{22}$ satisfy $\psi_{-}=-\arctan\left(\frac{Z_\mathrm{in}}{\eta}\frac{\alpha_{-}}{k}\right)$ and $\psi_{+}=-\arctan\left(\frac{Z_\mathrm{out}}{\eta}\frac{\alpha_{+}}{k}\right)$, respectively.

 \begin{figure}[!t]
 \centering
\includegraphics[width=8.5cm]{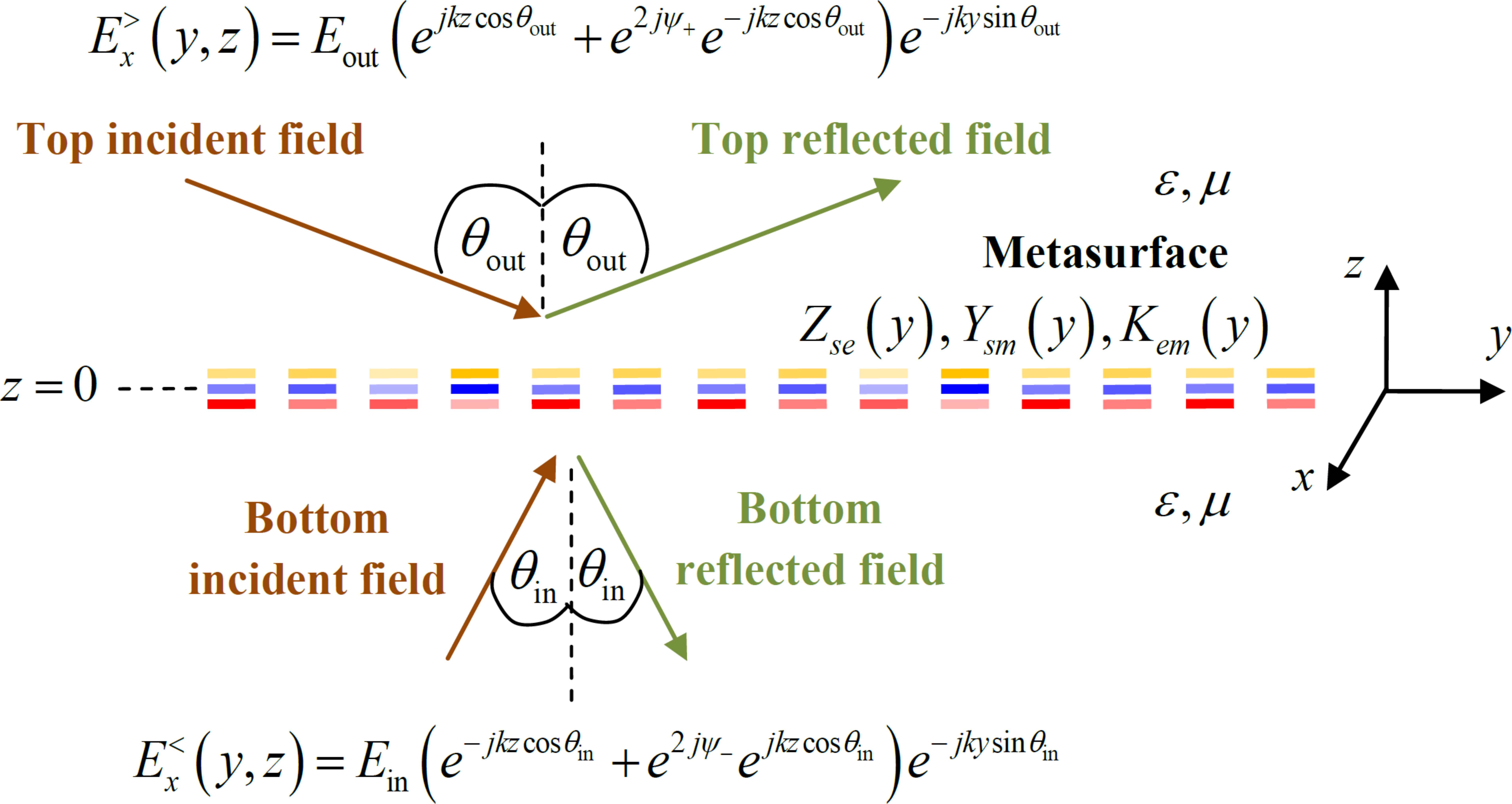}%
\caption{Physical configuration of an O-BMS supporting propagation of different surface waves on each of its facets (Fig. \ref{fig:physical_configuration_surface_waves}), when excited by plane waves. Upon plane-wave excitation these O-BMSs act as metamirrors, fully-reflecting the impinging power while the reflected fields incure a prescribed phase shift.}
\label{fig:physical_configuration_metamirror}
\end{figure} 

The matrix $\mathbf{[G]}$ of \eqref{equ:G_matrix_surface_waves} reveals another aspect of the surface-guiding O-BMSs given by \eqref{equ:OBMS_specifications_surface_waves}: these are actually metamirrors \cite{Radi2014_1,Asadchy2015}. Specifically, \eqref{equ:G_matrix_surface_waves} indicates that a plane wave incident on either side of the metasurface will be fully-reflected, incurring a phase shift of $2\psi_{-}$ or $2\psi_+$ depending whether the plane wave impinges on the bottom or top facet of the metasurface, respectively (Fig. \ref{fig:physical_configuration_metamirror}). The reflection phase for either facet is independent of the excitation from the other side, and is determined solely by the angle of incidence of the individual plane wave and the O-BMS properties. To further highlight this decoupling, we provide explicit expressions for such plane-wave fields supported by the O-BMS of \eqref{equ:OBMS_specifications_surface_waves}, namely,
\begin{align}
&\left\lbrace\!\!\!
\begin{array}{l}
	\vspace{3pt}
	E_x^<\left(y,z\right) = E_\mathrm{in}\left(\!\!\!\!\begin{array}{l}
	e^{-jk\cos\theta_\mathrm{in}z}\\ \,\,+G_{11}e^{jk\cos\theta_\mathrm{in}z}
	\end{array}\!\!\!\!\right)e^{-jk\sin\theta_\mathrm{in}y} \\
	\vspace{3pt}
	H_y^<\left(y,z\right) = \frac{1}{Z_\mathrm{in}}E_\mathrm{in}\left(\!\!\!\!\begin{array}{l}
	e^{-jk\cos\theta_\mathrm{in}z}\\ \,\,-G_{11}e^{jk\cos\theta_\mathrm{in}z}
	\end{array}\!\!\!\!\right)e^{-jk\sin\theta_\mathrm{in}y} 
\end{array}
\right. \nonumber \\
&\left\lbrace\!\!\!
\begin{array}{l}
	\vspace{3pt}
	E_x^>\left(y,z\right) = E_\mathrm{out}\left(\!\!\!\!\begin{array}{l}
	e^{jk\cos\theta_\mathrm{out}z}\\ \,\,+G_{22}e^{-jk\cos\theta_\mathrm{out}z}
	\end{array}\!\!\!\!\right)e^{-jk\sin\theta_\mathrm{out}y} \\
	\vspace{3pt}
	H_y^>\left(y,z\right) =-\frac{1}{Z_\mathrm{out}} E_\mathrm{out}\left(\!\!\!\!\begin{array}{l}
	e^{jk\cos\theta_\mathrm{out}z}\\ \,\,-G_{22}e^{-jk\cos\theta_\mathrm{out}z}
	\end{array}\!\!\!\!\right)e^{-jk\sin\theta_\mathrm{out}y},
\end{array}
\right.
\label{equ:fields_metamirror}
\end{align}
where the reflection coefficients $G_{11}$ and $G_{22}$ are given by \eqref{equ:G_matrix_surface_waves}, the wave impedances $Z_\mathrm{in}$ and $Z_\mathrm{out}$ are defined as in \eqref{equ:fields_plane_wave_refraction}, and we emphasize that the field amplitudes $E_\mathrm{in}$ and $E_\mathrm{out}$ can independently possess any arbitrary value.

As a final remark we note that although the reflection phase generally depends on the angle of incidence of the impinging plane wave, when $\alpha_{\pm}\gg k$ or $\alpha_{\pm}\ll k$ the reflection phase tends to $2\psi_{\pm}\rightarrow-\pi$ or $2\psi_{\pm}\rightarrow0$, respectively. This means that O-BMSs designed following \eqref{equ:OBMS_specifications_surface_waves} with very large or very small values of $\alpha_{\pm}/k$ will exhibit behaviour resembling perfect electric conductor (PEC) or perfect magnetic conductor (PMC), respectively, on the corresponding facet, with little sensitivity to the angle of incidence.

\subsection{Cavity-excited O-BMS antenna}
\label{subsec:cavity_excited_antenna}
\begin{figure*}[!b]
\normalsize
\setcounter{MYtempeqncnt}{\value{equation}}
\setcounter{equation}{19}
\hrulefill
\begin{equation}
\left\lbrace\!\!\!
\begin{array}{l}
	\vspace{3pt}	
	E_x^<\left(y,z\right) = k\eta \dfrac{{{I}_{0}}}{L} 
	\sum\limits_{m=0}^{\infty }\dfrac{1}{{{\beta }_{2m+1}}}
	\dfrac{e^{-j\beta_{2m+1}\left(z_<+d\right)}-e^{j\beta_{2m+1}\left(z_<+d\right)}}
	{e^{j\beta_{2m+1}d}-\Gamma_{2m+1}e^{-j\beta_{2m+1}d}}
	\left(e^{-j\beta_{2m+1}z_>}-\Gamma_{2m+1}e^{j\beta_{2m+1}z_>}\right)\cos \left( {{k}_{t,2m+1}}y \right)	
	 \\
	\vspace{3pt}
	H_y^<\left(y,z\right) = \dfrac{{{I}_{0}}}{L} 
	\sum\limits_{m=0}^{\infty }
	\dfrac{e^{-j\beta_{2m+1}\left(z_<+d\right)}\mp e^{j\beta_{2m+1}\left(z_<+d\right)}}
	{e^{j\beta_{2m+1}d}-\Gamma_{2m+1}e^{-j\beta_{2m+1}d}}
	\left(e^{-j\beta_{2m+1}z_>}\pm\Gamma_{2m+1}e^{j\beta_{2m+1}z_>}\right)\cos \left( {{k}_{t,2m+1}}y \right),
\end{array}
\right.
\label{equ:fields_antenna_below}
\end{equation}
\setcounter{equation}{20}
\end{figure*}
\begin{figure}[!t]
 \centering
\includegraphics[width=8.5cm]{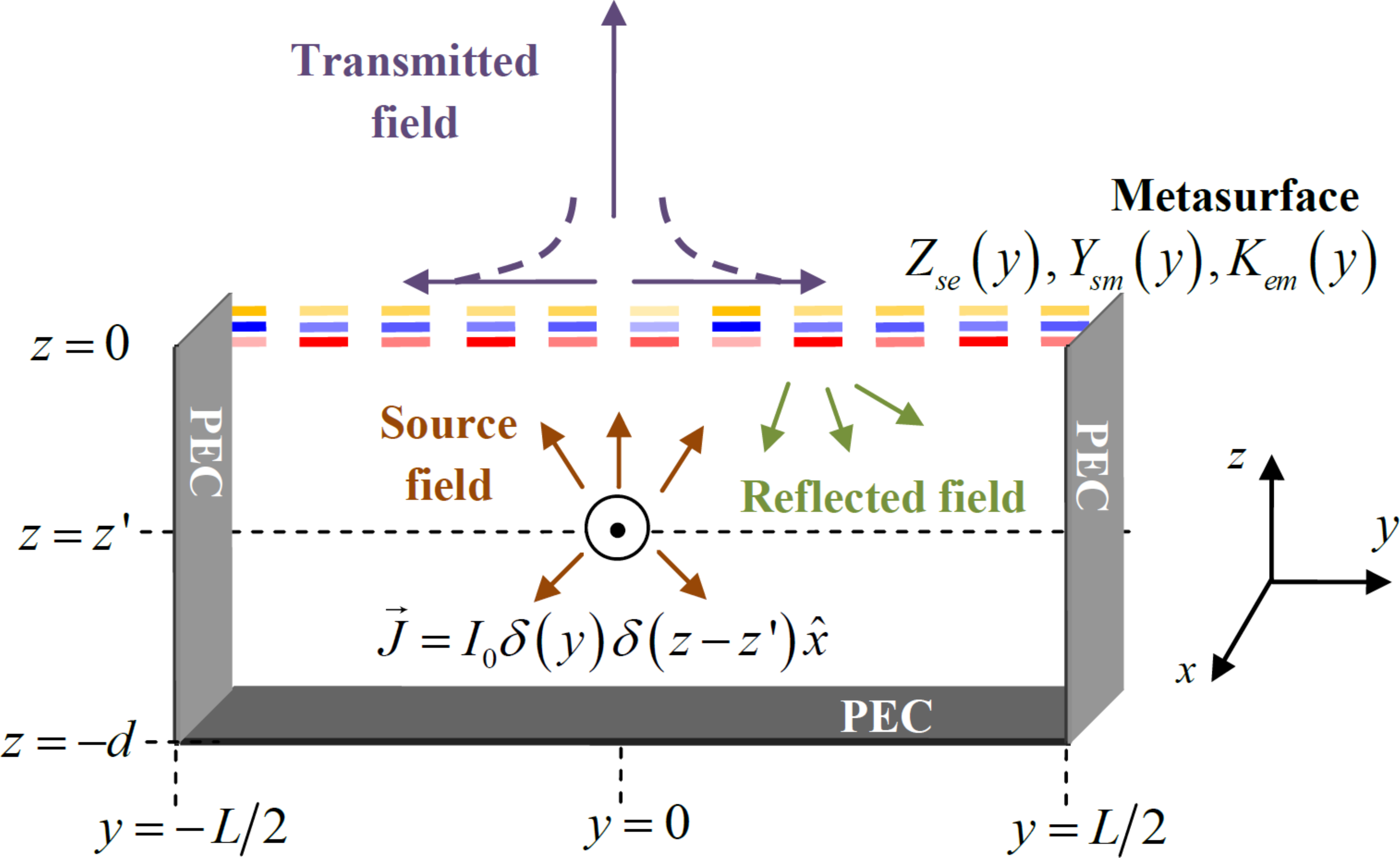}%
\caption{Physical configuration of a cavity-excited O-BMS antenna. The electric line source excitation is coupled predominantly to the highest-order fast mode of the lateral cavity, yielding uniform illumination of the metasurface. The O-BMS is engineered to induce the suitable currents to form a highly-directive broadside beam accompanied by a "standing" surface wave on the aperture.}
\label{fig:physical_configuration_antenna}
\end{figure}

The final application we consider herein utilizes the O-BMS to design a highly-directive low-profile antenna excited by a simple single source. To that end, we consider the configuration depicted in Fig. \ref{fig:physical_configuration_antenna}, where an electric line source is positioned at $\left(y,z\right)=\left(0,z'\right)$ below the metasurface ($z'<0$), surrounded by three PEC walls located at $z=-d$, $y=\pm L/2$; as before, the O-BMS (of length $L$) is situated at $z=0$.

Similar configuration using a HMS instead of an O-BMS was utilized recently to demonstrate antennas exhibiting near-unity aperture illumination efficiencies from arbitrarily large apertures \cite{Epstein2016}. The basic idea was to optimize the cavity excitation such that the highest-order fast lateral mode is predominantly excited, ensuring the aperture is well illuminated, and design the HMS following \cite{Epstein2014} to maintain aperture fields with uniform phase. \textcolor{black}{As discussed therein, HMSs can be designed to generate highly-directive radiation utilizing an arbitrary source configuration with passive and lossless components, however they must adhere to the local impedance equalization and local power conservation conditions. The former condition determines the reflection coefficient in the spectral domain, coercing cavity fields which consist of a superposition of lateral modes. Although the cavity mode composition can be optimized to achieve very high aperture illumination efficiencies, the constraint on the reflection coefficients places a limit to the achievable performance due to the unavoidable multimode excitation.}

\textcolor{black}{On the other hand, as proved in Subsection \ref{subsec:macro_design} and demonstrated in Subsections \ref{subsec:refraction} and \ref{subsec:surface_wave}, for a passive lossless O-BMS-based device, the only constraint applicable is the local power conservation \eqref{equ:local_power_conservation}. The practical implication of this result is that the additional magnetoelectric degree of freedom associated with O-BMSs releases the constraint on the metasurface reflection coefficient, allowing us to engineer its value (in the spectral domain) at will. Indeed, in this Subsection, we harness this additional degree of freedom to guarantee that only a single (favourable) cavity mode carries real power to the metasurface plane, achieving an optimal $100\%$ aperture illumination efficiency with only half the HMS-based antenna thickness.}

As in the previous subsections, our goal is to stipulate the fields below and above the metasurface such that they obey Maxwell's equations including the relevant boundary conditions, source conditions, and radiation conditions; local power conservation \eqref{equ:local_power_conservation} is satisfied; and the desirable functionality is achieved. Once all three conditions are met, we invoke \eqref{equ:passive_lossless_design} to obtain the required (passive and lossless) O-BMS design. 

We therefore begin by writing a general expression for the fields inside the cavity (i.e. below the metasurface $z<0$). For an electric line source carrying a current of $\vec{J}=I_0\delta\left(y\right)\delta\left(z-z'\right)\hat{x}$ these are given by \eqref{equ:fields_antenna_below} at the bottom of this page, 
where we define $z_>\triangleq\max\left\{z,z'\right\}$ and $z_<\triangleq\min\left\{z,z'\right\}$; the upper (lower) sign corresponds to the field at $z>z'$ ($z<z'$), facilitating the discontinuity in the magnetic field due to the localized source \cite{Epstein2016, FelsenMarcuvitz1973}. The $m$th term in the summations is associated with the $n$th lateral mode of the cavity ($n=2m+1$), characterized by a transverse wavenumber $k_{t,n}=n\pi/L$ and a longitudinal wavenumber $\beta_n=\sqrt{k^2-k_{t,n}^2}$; the respective metasurface reflection coefficient is denoted by $\Gamma_{n}$. The numerator in \eqref{equ:fields_antenna_below} corresponds to the interference between the source and its image, induced by the back-PEC at $z=-d$, whereas the denominator manifests multiple reflections between the metasurface and this back-PEC.

In \cite{Epstein2016} we have shown that the highest-order fast mode forms a favourable aperture illumination for achieving highly-directive radiation. The power profile associated with this mode forms hot spots on the metasurface aperture which are approximately half a wavelength apart. From antenna array theory it is known that linearly-phased current excitations with such a separation distance yield very directive radiation without grating lobes, regardless of the scan angle. 

Therefore, we optimize the metasurface reflection coefficient and cavity configuration to obtain exclusive excitation of the highest-order fast mode on the aperture, i.e. the $n=2N-1$ mode for an aperture of $L=N\lambda$ ($N\in\mathbb{N}$). This is achieved by setting the cavity thickness $d$ such that
\begin{equation}
	\beta_{2N-1}d = \dfrac{\pi}{2} \Rightarrow d=\dfrac{\lambda}{4}\dfrac{2N}{\sqrt{4N-1}},
\label{equ:cavity_thickness}
\end{equation}  
and the O-BMS reflection coefficient such that
\begin{equation}
	\Gamma_n = \left\{
	\begin{array}{l l}
		-\left|\Gamma_{2N-1}\right| & n=2N-1 \\
		-1 & n\neq 2N-1,
	\end{array}
	\right.
\label{equ:cavity_OBMS_reflection_coefficient}
\end{equation} 
where the reflection coefficient of the $\left(2N-1\right)$th mode can be chosen arbitrarily as long as $\left|\Gamma_{2N-1}\right|<1$. \textcolor{black}{The cavity thickness and the reflection coefficients were chosen such that only the highest-order fast mode could contribute to the real power crossing the metasurface, whereas the other modes yield purely-reactive power. More specifically, with these settings, the tangential fields just below the metasurface are given by (note the $\pi/2$ phase shift between the first and second terms in the electric field expression)}
\begin{equation}
\left\lbrace\!\!\!
\begin{array}{l}
	\vspace{3pt}	
	\begin{array}{l}
	\vspace{3pt}
	E_x^-\left(y\right) = -2k\eta \frac{{{I}_{0}}}{L}\frac{\sin\left[\beta_\mathrm{2N-1}\left(z'+d\right)\right]}{\beta_{2N-1}}\frac{1+\left|\Gamma_{2N-1}\right|}{1-\left|\Gamma_{2N-1}\right|} \cos \left( {{k}_{t,2N-1}}y \right) \\
	\,\, -2jk\eta\frac{I_0}{L}\sum_{m\neq N-1}^{\infty} \frac{1}{\beta_{2m+1}}\frac{\sin\left[\beta_{2m+1}\left(z'+d\right)\right]}{\cos\left(\beta_{2m+1}d\right)}\cos \left( {{k}_{t,2m+1}}y \right)		
	\end{array}
	 \\
	\vspace{3pt}
	H_y^-\left(y\right) = -2\frac{{{I}_{0}}}{L} 
	\sin\left[\beta_{2N-1}\left(z'+d\right)\right]\cos\left(k_{t,2N-1}y\right),
\end{array}
\right.
\label{equ:fields_antenna_below_atOBMS}
\end{equation}
and the corresponding real power incident the O-BMS from below is
\begin{equation}
\begin{array}{l}
P_z^{-}\left(y\right)= \\
\,\, =\textstyle 2k\eta\left|\frac{I_0}{L}\right|^2\frac{\sin^2\left[\beta_{2N-1}\left(z'+d\right)\right]}{\beta_{2N-1}}\frac{1+\left|\Gamma_{2N-1}\right|}{1-\left|\Gamma_{2N-1}\right|}\cos^2\left(k_{t,2N-1}y\right).
\end{array}
\label{equ:real_power_antenna_below_atOBMS}
\end{equation}

This last result is actually very compelling as the same power profile could be matched to aperture fields (on the upper facet of the metasurface) which produce highly-directive broadside radiation. In other words, stipulating the fields above the metasurface ($z>0$) to follow
\begin{equation}
\left\lbrace\!\!\!
\begin{array}{l}
	\vspace{3pt}
	E_x^>\left(y,z\right) = E_\mathrm{out}\left[e^{-jkz}+e^{\Im\left\{\beta_{nr}\right\}z}\cos\left(2k_{t,2N-1}y\right)\right] \\
	\vspace{3pt}
	H_y^>\left(y,z\right) =\frac{E_\mathrm{out}}{\eta} 
	\left[
	\begin{array}{l}
	e^{-jkz} \\
	+j\frac{\Im\left\{\beta_{nr}\right\}}{k}e^{\Im\left\{\beta_{nr}\right\}z}\cos\left(2k_{t,2N-1}y\right)
	\end{array}
	\right],
\end{array}
\right.
\label{equ:fields_antenna_above}
\end{equation}
where the non-radiative longitudinal wavenumber is $\beta_{nr}=-j\sqrt{\left(2k_{t,2N-1}\right)^2-k^2}$, would yield a (real) power profile $P_z^+\left(y\right)$ identical to the one in \eqref{equ:real_power_antenna_below_atOBMS} if
\begin{equation}
E_\mathrm{out} = \textstyle \eta\left|\frac{I_0}{L}\right|\sqrt{\frac{2\beta_{2N-1}}{k}\frac{1+\left|\Gamma_{2N-1}\right|}{1-\left|\Gamma_{2N-1}\right|}}
	\sin\left[\beta_{2N-1}\left(z'+d\right)\right]e^{-j\xi_\mathrm{out}},
\label{equ:fields_antenna_above_amplitude}
\end{equation}
where $\xi_\mathrm{out}$ is an arbitrary uniform phase shift that can be imposed on the transmitted fields.

\textcolor{black}{It should be noted that although the fields above the metasurface \eqref{equ:fields_antenna_above} are defined as if the metasurface is of infinite extent, they are in fact supported by an O-BMS of a finite length $L$, and should be truncated correspondingly. Our implicit assumption here, similar to \cite{Epstein2014,Epstein2016}, is that the aperture is large enough such that the fields just above the metasurface do not differ significantly from the ones evaluated by substituting $z\rightarrow 0^+$ into \eqref{equ:fields_antenna_above} [which are used to establish the local power conservation via \eqref{equ:fields_antenna_above_amplitude}]. Subsequently, to evaluate the antenna radiation pattern, we assume that these aperture fields (i.e., at $z\rightarrow 0^+$) follow \eqref{equ:fields_antenna_above} for $y<\left|L/2\right|$ and vanish outside the metasurface aperture $y>\left|L/2\right|$, and use standard asymptotic techniques (\textit{cf.} Appendix C of \cite{Epstein2014}) to propagate them to the far field. This approximation is valid if the aperture is sufficiently long such that edge effects are negligible, and should be verified with full-wave simulations of the finite structure. 
}

We utilize this formulation to design a highly-directive cavity-excited O-BMS antenna with an aperture of $L=10\lambda$ ($2N-1=19$). Following \eqref{equ:cavity_thickness}, we set the cavity thickness to $d=0.801\lambda$. As it turns out that for this thickness, the $7$th lateral mode (and not only the $19$th mode) also satisfies $\beta_{7}d=\pi/2$, we must set the source position $z'$ to suppress the excitation of this mode [\eqref{equ:fields_antenna_below_atOBMS} is derived under the assumption that only the $\left(2N-1\right)$th mode satisfies the constructive interference condition formulated by \eqref{equ:cavity_thickness}]. Consequently, the source position is set such that $\beta_7\left(z'+d\right)=\pi$, guaranteeing that the terms corresponding to the $7$th mode will vanish in \eqref{equ:fields_antenna_below} due to destructive interference between the source and its image; specifically, $z'=-0.267\lambda$. To reduce the quality factor of the cavity as much as possible, we set the reflection coefficient of the $\left(2N-1\right)$th mode  \eqref{equ:cavity_OBMS_reflection_coefficient} to $\Gamma_{2N-1}=0$ (a higher value was explored in \cite{Epstein2016_1}). This fixes the last degree of freedom in the design, and the required O-BMS constituents can be evaluated by substituting \eqref{equ:fields_antenna_below_atOBMS} and \eqref{equ:fields_antenna_above}, with \eqref{equ:fields_antenna_above_amplitude} and the prescribed values of $d$, $z'$, and $\Gamma_{2N-1}$, into \eqref{equ:passive_lossless_design}.

\begin{figure}[htbp]
 \centering
\includegraphics[width=6cm]{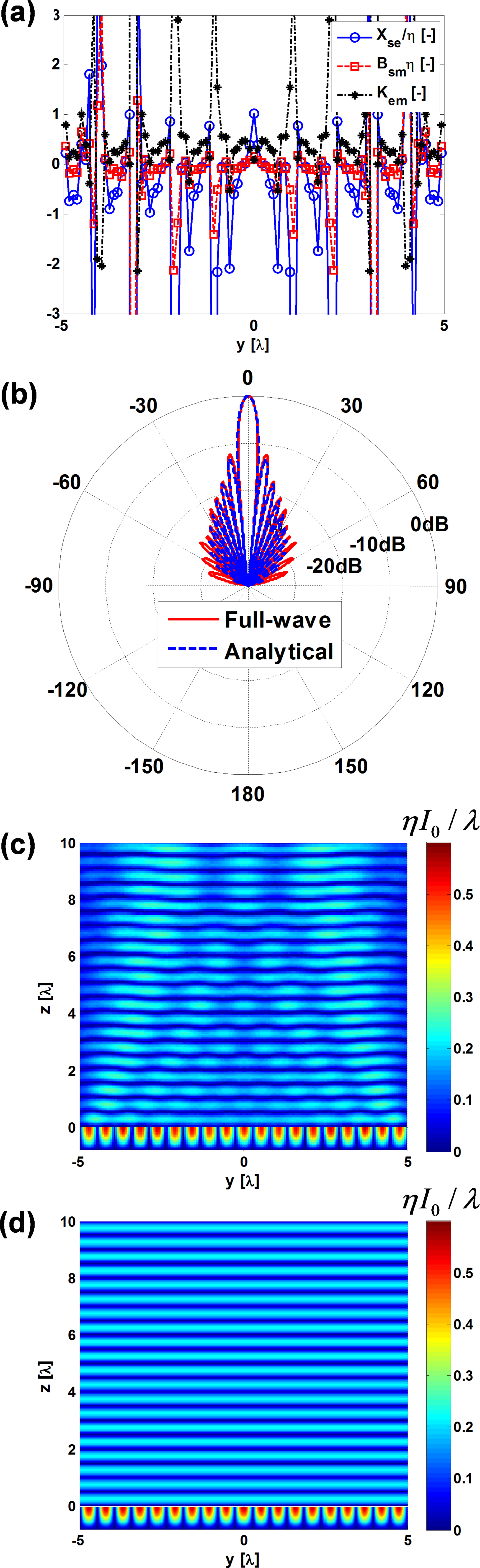}%
\caption{Cavity-excited O-BMS antenna of aperture length $L=10\lambda$ ($\xi_\mathrm{out}=90^{\circ}$). (a) Required electric surface reactance $X_{se}$ (blue solid lines), magnetic surface susceptance $B_{sm}$ (red dashed lines), and magnetoelectric coupling coefficient $K_{em}$ (black dash-dotted lines). (b) Normalized radiation pattern as recorded in full-wave simulations (solid red line) and as predicted by the analytical formulas (blue dashed line); the latter is obtained using the aperture fields derived from \eqref{equ:fields_antenna_above} and the scheme presented in \cite{Balanis1997}. (c) Simulated electric field distribution $\left|\Re\left\{E_x\left(y,z\right)\right\}\right|$. (d) Analytical prediction of $\left|\Re\left\{E_x\left(y,z\right)\right\}\right|$ via \eqref{equ:fields_antenna_below} and \eqref{equ:fields_antenna_above}.}
\label{fig:results_antenna}
\end{figure}

The resultant O-BMS specifications, presented in Fig. \ref{fig:results_antenna}(a), were used to implement the antenna device according to the microscopic design procedure outlined in Subsection \ref{subsec:micro_design}. The radiation patterns as obtained from full-wave simulations and analytical formulas presented in Fig. \ref{fig:results_antenna}(b) are in excellent agreement. The excitation of the highest-order fast lateral mode is clearly observed in the simulated field snapshots [Fig. \ref{fig:results_antenna}(c)] and the ones evaluated analytically [Fig. \ref{fig:results_antenna}(d)], leading to very high aperture illumination efficiencies. This is confirmed by the approximately-parallel phase fronts covering the entire aperture length, indicating high phase purity. The discrepancies observed in the full-wave simulation results of Fig. \ref{fig:results_antenna}(c) are attributed to the minor spatial dispersion exhibited by the meta-atoms \textcolor{black}{and the finite length of the metasurface [see our comment after \eqref{equ:fields_antenna_above_amplitude}]}. 
Nevertheless, the radiation characteristics of the antennas, summarized in Table \ref{tab:antenna_performance} along with the parameters associated with a uniformly-illuminated aperture of the same length \cite{Balanis1997}, indicate a very good agreement between full-wave simulations and analytical predictions. \textcolor{black}{In fact, comparison with the performance of a uniformly-illuminated aperture reveals that the aperture illumination efficiency, defined as the ratio between the antenna directivity and the one corresponding to a uniform radiating aperture, practically coincides with the ideal $100\%$, both for the simulated antenna and in analytical predictions.}

This example highlights several powerful features of O-BMS-based devices designed following the presented method. First, as demonstrated by \eqref{equ:cavity_OBMS_reflection_coefficient}, the O-BMS reflection can be meticulously engineered, enabling control of the reflection coefficient (magnitude and phase) of \emph{individual} spectral components. Second, as exemplified by the rather complex cavity excitation \eqref{equ:fields_antenna_below}, our general design procedure can be readily applied to highly-intricate configurations, which allow incorporation of a variety of sources and scatterers, and offer a multitude of degrees of freedom. Third, as the only preliminary requirement for using the design scheme of \eqref{equ:passive_lossless_design} is local conservation of the \emph{real} power, reactive fields can be harnessed as an additional degree of freedom to establish field transformations which \emph{rigorously} comply with the design procedure.

\begin{table}[!t]
\centering
\begin{threeparttable}[b]
\renewcommand{\arraystretch}{1.3}
\caption{Radiation characteristics of cavity-excited O-BMS antennas (corresponding to Fig. \ref{fig:results_antenna}).}
\label{tab:antenna_performance}
\centering
\footnotesize{
\begin{tabular}{l|c c c}
\hline \hline
& Full-wave & Analytical & Uniform  \\
\hline \hline \\[-1.3em]	
	 \begin{tabular}{l} Half-power beamwidth \end{tabular}
	 & $4.83^{\circ}$ & $5.07^{\circ}$ & $5.08^{\circ}$	 
	 	 	\\	\hline 
	 \begin{tabular}{l} Directivity (2D) [dBi] \end{tabular}
	 & $18.09$ & $18.03$ & \textcolor{black}{$18.05$}	 
	 	 	 	 \\	\hline 
	 \begin{tabular}{l} First Side-Lobe \end{tabular}
	 & $8.1^{\circ}$ & $8.2^{\circ}$ & $8.2^{\circ}$	 
	 	 	 	 \\	\hline 
	 \begin{tabular}{l} Side-Lobe Level [dB] \end{tabular}
	 & $-12.1$ & $-13.3$ & $-13.3$
	 	 	 	 \\ 	
\hline \hline
\end{tabular}}
\end{threeparttable}
\end{table}

\textcolor{black}{It should be emphasized that this is \emph{not} an analysis problem, attempting to resolve the fields scattered by a given inhomogeneous metasurface covering a cavity, which may be difficult to tackle analytically using scalar spectrally-decoupled reflection coefficients. Within the framework of our design procedure, we can stipulate \emph{arbitrary} fields below ($z<0$) and above ($z>0$) the metasurface provided that they adhere to Maxwell's equations (including the suitable source, radiation, and boundary conditions), and that they satisfy local power conservation. As the modal expressions \eqref{equ:fields_antenna_below} represent such proper Maxwellian solutions for \emph{arbitrary} values of $d$ and $\Gamma_n$, we can tune these parameters freely (as long as the corresponding power profile is locally matched by the transmitted fields). Therefore, even if these field expressions relate to only a subset of possible metasurface designs (the ones that can be solved using modal analysis), they are perfectly valid for use via our synthesis scheme, and the resulting metasurface specifications will support them. 
}

Indeed, utilizing these features for the design yields a low-profile cavity-excited O-BMS antenna with a \textcolor{black}{perfect aperture illumination efficiency}. The ability to engineer the spectral response of the O-BMS reflection coefficient is harnessed to guarantee that only the highest-order fast mode is expressed in the real power exciting the metasurface \eqref{equ:real_power_antenna_below_atOBMS}, establishing a regular and uniform array-like illumination of the aperture, regardless of the aperture size. The possibility to integrate the probe-fed cavity configuration into the design procedure in a straightforward manner provides additional degrees of freedom to ensure this mode purity, while at the same time mitigates edge-taper losses. Finally, by using a non-radiative "standing" surface wave (with transverse wavenumber $2k_{t,2N-1}$ and longitudinal wavenumber $\beta_{nr}$) in addition to the desirable broadside-radiating aperture fields \eqref{equ:fields_antenna_above_amplitude}, we were able to \emph{accurately} match the BSTCs \eqref{equ:BSTCs} with a passive and lossless O-BMS specifications. 
This outperforms our previously-introduced cavity-excited HMS antenna \cite{Epstein2016}, guaranteeing optimal aperture illumination efficiency with half the device thickness.

\section{Conclusion}
\label{sec:conclusion}
To conclude, we have presented a general theory for the design of passive lossless omega-type bianisotropic metasurfaces, implementing a given (arbitrary) field transformation. Our only requirement from this transformation is that it would locally conserve the real power crossing the metasurface at each point. Once the desirable (Maxwellian) fields just below and above the metasurface are stipulated, the derived analytical formulas can be used to evaluate the (macroscopic) O-BMS constituents. Importantly, by utilizing a two-port microwave network circuit model for an O-BMS unit cell we have shown that the microscopic implementation of each specified omega-type meta-atom could be achieved via asymmetric cascade of three impedance sheets, whose surface impedance values can also be assessed analytically. The combined macroscopic and microscopic design procedures allow systematic engineering of O-BMS-based devices for a wide range of applications. Formulated as a natural generalization of the analogous methodologies derived previously for Huygens' metasurfaces \cite{Pfeiffer2013, Monticone2013, Selvanayagam2013_2, Epstein2015_3}, the complete derivation highlights the role of the bianisotropy in establishing a given field transformation.

As opposed to standard HMSs, which can generally control the phase of transmitted fields, we have shown that O-BMSs can be used to engineer the reflection coefficient phase and amplitude as well. The origin of this fundamental difference is observed when the plane-wave refraction functionality is examined: while the symmetric HMSs can only exhibit the same input (wave) impedance to both its facets \cite{Epstein2014_2}, the asymmetry inherent to O-BMSs allows engineering of the input impedance on each facet independently \cite{Wong2015_1}. 

Subsequently, we have demonstrated the power and versatility of the proposed scheme by designing three different metasurface devices, involving diverse reflection and transmission coefficient control and a variety of excitation sources. Reflectionless wide-angle refracting O-BMSs achieve zero reflection in the face of extreme wave impedance mismatch; independent surface-wave guiding O-BMSs are based on a fully-reflective asymmetric structure (metamirror); and the cavity-excited O-BMS antennas which employ reflection coefficient control of individual spectral components to achieve unity aperture illumination efficiencies with probe-fed low-profile large-aperture radiators. 

These examples verify our theoretical derivation, indicating an excellent agreement between analytical predictions and full-wave simulations. In addition, they point out the immense potential of O-BMSs for a wide variety of applications, incorporating complex scatterers and sources and high-level reflection coefficient control to achieve performance and functionalities not achievable with standard HMSs. Therefore, we believe that the presented design procedure, systematically addressing both macroscopic and microscopic aspects, may serve as a primary tool for engineering of next-generation metasurface-based devices.

\appendices
\textcolor{black}{\section{Derivation of the passive lossless O-BMS design specifications \eqref{equ:passive_lossless_design}}\label{app:OBMS_design_formulas}}

\textcolor{black}{For completeness, we provide in this appendix a detailed derivation of \eqref{equ:passive_lossless_design}. We thus consider a 2D configuration with TE-polarized fields as described in Section \ref{subsec:macro_design} (Fig. \ref{fig:physical_configuration}), with the fields above ($z>0$) and below ($z<0$) the metasurface given by $\left\{E_x^>\left(y,z\right),H_y^>\left(y,z\right),H_z^>\left(y,z\right)\right\}$ and $\left\{E_x^<\left(y,z\right),H_y^<\left(y,z\right),H_z^<\left(y,z\right)\right\}$, respectively. We assume the stipulated fields on the top ($z\rightarrow 0^+$) and bottom ($z\rightarrow 0^-$) facets of the metasurface satisfy local power conservation \eqref{equ:local_power_conservation}; by definition, the relation between these fields is given by the O-BSTCs \eqref{equ:BSTCs}. For such a given set of fields, the O-BSTCs form two complex equations with three complex unknowns $Z_{se}, Y_{sm}, K_{em}$; our goal is to find a solution to this set of equations which corresponds to a passive and lossless design, i.e. where $\Re\{Z_{se}\}=\Re\{Y_{sm}\}=\Im\{K_{em}\}=0$.
}

\textcolor{black}{The first step in the derivation includes multiplication of the first and second equations in \eqref{equ:BSTCs} by $\left(H_y^+ - H_y^-\right)^*$ and $\left(E_x^+ - E_x^-\right)^*$, respectively. Equating the real and imaginary parts of the resulting two \emph{complex} equations, and expressing the unknowns via their real and imaginary components, yields the following set of four \emph{real} equations (featuring six \emph{real} unknowns, namely, $\Re\{Z_{se}\},\Im\{Z_{se}\},\Re\{Y_{sm}\},\Im\{Y_{sm}\},\Re\{K_{em}\},\Im\{K_{em}\}$):
\begin{equation}
\left\{\!\!\!
	\begin{array}{l}
	\Re\left\{\left(E_{x}^{+}+E_{x}^{-}\right)\left( H_{y}^{+}-H_{y}^{-} \right)^*\right\} =  -2\Re\left\{{Z}_{se}\right\}\left| H_{y}^{+}-H_{y}^{-} \right|^2 \\
	\quad\quad -2\Re\left\{{K}_{em}\right\}\Re\left\{\left(E_{x}^{+}-E_{x}^{-}\right)\left( H_{y}^{+}-H_{y}^{-} \right)^*\right\}  \\
	\quad\quad +2\Im\left\{{K}_{em}\right\}\Im\left\{\left(E_{x}^{+}-E_{x}^{-}\right)\left( H_{y}^{+}-H_{y}^{-} \right)^*\right\}
	\\
   \Re\left\{\left(H_{y}^{+}+H_{y}^{-}\right)\left( E_{x}^{+}-E_{x}^{-} \right)^*\right\}=-2\Re\left\{{Y}_{sm}\right\}\left| E_{x}^{+}-E_{x}^{-} \right|^2 \\
   \quad\quad +2\Re\left\{{K}_{em}\right\}\Re\left\{\left(H_{y}^{+}-H_{y}^{-}\right)\left( E_{x}^{+}-E_{x}^{-} \right)^*\right\} \\
   \quad\quad -2\Im\left\{{K}_{em}\right\}\Im\left\{\left(H_{y}^{+}-H_{y}^{-}\right)\left( E_{x}^{+}-E_{x}^{-} \right)^*\right\} \\
	\Im\left\{\left(E_{x}^{+}+E_{x}^{-}\right)\left( H_{y}^{+}-H_{y}^{-} \right)^*\right\} =  -2\Im\left\{{Z}_{se}\right\}\left| H_{y}^{+}-H_{y}^{-} \right|^2 \\
	\quad\quad -2\Im\left\{{K}_{em}\right\}\Re\left\{\left(E_{x}^{+}-E_{x}^{-}\right)\left( H_{y}^{+}-H_{y}^{-} \right)^*\right\}  \\
	\quad\quad -2\Re\left\{{K}_{em}\right\}\Im\left\{\left(E_{x}^{+}-E_{x}^{-}\right)\left( H_{y}^{+}-H_{y}^{-} \right)^*\right\}
	\\
   \Im\left\{\left(H_{y}^{+}+H_{y}^{-}\right)\left( E_{x}^{+}-E_{x}^{-} \right)^*\right\}=-2\Im\left\{{Y}_{sm}\right\}\left| E_{x}^{+}-E_{x}^{-} \right|^2 \\
   \quad\quad +2\Im\left\{{K}_{em}\right\}\Re\left\{\left(H_{y}^{+}-H_{y}^{-}\right)\left( E_{x}^{+}-E_{x}^{-} \right)^*\right\} \\
   \quad\quad +2\Re\left\{{K}_{em}\right\}\Im\left\{\left(H_{y}^{+}-H_{y}^{-}\right)\left( E_{x}^{+}-E_{x}^{-} \right)^*\right\}.
	\end{array}
\right.
\label{equ:real_BSTCs}
\end{equation}
}

\textcolor{black}{Equation \eqref{equ:real_BSTCs} forms \emph{four} linear equations with \emph{six} unknowns, implying that \emph{two} degrees of freedom are redundant. Therefore, to promote passive and lossless solutions, we set two unknowns following
\begin{equation}
\Re\left\{Z_{se}\right\}=\Re\left\{Y_{sm}\right\}=0
\label{equ:passive_lossless_Zse_Ysm}
\end{equation}
and attempt to derive a consistent valid solution to \eqref{equ:real_BSTCs}.
}

\textcolor{black}{With this choice, the first two equations of \eqref{equ:real_BSTCs} become
\begin{equation}
\left\{\!\!\!
	\begin{array}{l}
	\Re\left\{\left(E_{x}^{+}+E_{x}^{-}\right)\left( H_{y}^{+}-H_{y}^{-} \right)^*\right\} = \\ 
	\quad\quad -2\Re\left\{{K}_{em}\right\}\Re\left\{\left(E_{x}^{+}-E_{x}^{-}\right)\left( H_{y}^{+}-H_{y}^{-} \right)^*\right\}  \\
	\quad\quad +2\Im\left\{{K}_{em}\right\}\Im\left\{\left(E_{x}^{+}-E_{x}^{-}\right)\left( H_{y}^{+}-H_{y}^{-} \right)^*\right\}
	\\
   \Re\left\{\left( E_{x}^{+}-E_{x}^{-} \right)\left(H_{y}^{+}+H_{y}^{-}\right)^*\right\}= \\
   \quad\quad +2\Re\left\{{K}_{em}\right\}\Re\left\{\left( E_{x}^{+}-E_{x}^{-} \right)\left(H_{y}^{+}-H_{y}^{-}\right)^*\right\} \\
   \quad\quad +2\Im\left\{{K}_{em}\right\}\Im\left\{\left( E_{x}^{+}-E_{x}^{-} \right)\left(H_{y}^{+}-H_{y}^{-}\right)^*\right\},
	\end{array}
\right.
\label{equ:real_BSTCs_passive_lossless_Zse_Ysm}
\end{equation}
where we used the complex-number identities $\Re\left\{w\right\}=\Re\left\{w^*\right\}$ and $\Im\left\{w\right\}=-\Im\left\{w^*\right\}$, valid for every $w\in\mathbb{C}$.
}

\textcolor{black}{Summing the two equations of \eqref{equ:real_BSTCs_passive_lossless_Zse_Ysm} and using local power conservation \eqref{equ:local_power_conservation} leads to
\begin{equation}
\Im\left\{{K}_{em}\right\}\Im\left\{\left(E_{x}^{+}-E_{x}^{-}\right)\left( H_{y}^{+}-H_{y}^{-} \right)^*\right\}=0,
\label{equ:passive_lossless_Kem}
\end{equation}
which is solved by stipulating a passive lossless magnetoelectric coefficient, i.e.
\begin{equation}
\Im\left\{{K}_{em}\right\}=0.
\label{equ:passive_lossless_Kem_explicit}
\end{equation}
Subtracting the two equations of \eqref{equ:real_BSTCs_passive_lossless_Zse_Ysm}, and using \eqref{equ:passive_lossless_Kem_explicit}, yields the desirable closed-form expression for $K_{em}\in\mathbb{R}$ presented in \eqref{equ:passive_lossless_design}. 
}

\textcolor{black}{Equations \eqref{equ:passive_lossless_Zse_Ysm} and \eqref{equ:passive_lossless_Kem_explicit} imply that the solution achieved so far is indeed passive and lossless, i.e. $K_{em}$ is real whereas $Z_{se}$ and $Y_{sm}$ are purely imaginary. Substituting these results into the \emph{last} two equations of \eqref{equ:real_BSTCs} yields
\begin{equation}
\left\{\!\!\!
	\begin{array}{l}
	j{Z}_{se}\left| H_{y}^{+}-H_{y}^{-} \right|^2=\frac{1}{2}\Im\left\{\left(E_{x}^{+}+E_{x}^{-}\right)\left( H_{y}^{+}-H_{y}^{-} \right)^*\right\} \\	
	\quad\quad +{K}_{em}\Im\left\{\left(E_{x}^{+}-E_{x}^{-}\right)\left( H_{y}^{+}-H_{y}^{-} \right)^*\right\}
	\\
	\vspace{3pt}
 j{Y}_{sm}\left| E_{x}^{+}-E_{x}^{-} \right|^2=\frac{1}{2}\Im\left\{\left(H_{y}^{+}+H_{y}^{-}\right)\left( E_{x}^{+}-E_{x}^{-} \right)^*\right\} \\   
   \quad\quad -{K}_{em}\Im\left\{\left(H_{y}^{+}-H_{y}^{-}\right)\left( E_{x}^{+}-E_{x}^{-} \right)^*\right\}.
	\end{array}
\right.
\label{equ:real_BSTCs_passive_lossless_all}
\end{equation}
Dividing the first equation by $\left| H_{y}^{+}-H_{y}^{-} \right|^2$ and the second equation by $\left| E_{x}^{+}-E_{x}^{-} \right|^2$, which are real quantities, leads to the desirable expressions for $Z_{se}$ and $Y_{sm}$, respectively, as presented in \eqref{equ:passive_lossless_design}.
}

\textcolor{black}{As a final step, the expressions presented in \eqref{equ:passive_lossless_design} can be substituted into \eqref{equ:BSTCs} to verify that a tautology is formed, provided that local power conservation \eqref{equ:local_power_conservation} is satisfied by the field quantities. 
}

\textcolor{black}{\section{Physical structure for three-layer O-BMS meta-atoms}\label{app:meta_atom_physical_structure}}

\begin{figure}[!b]
\textcolor{black}{
\centering
\includegraphics[width=6cm]{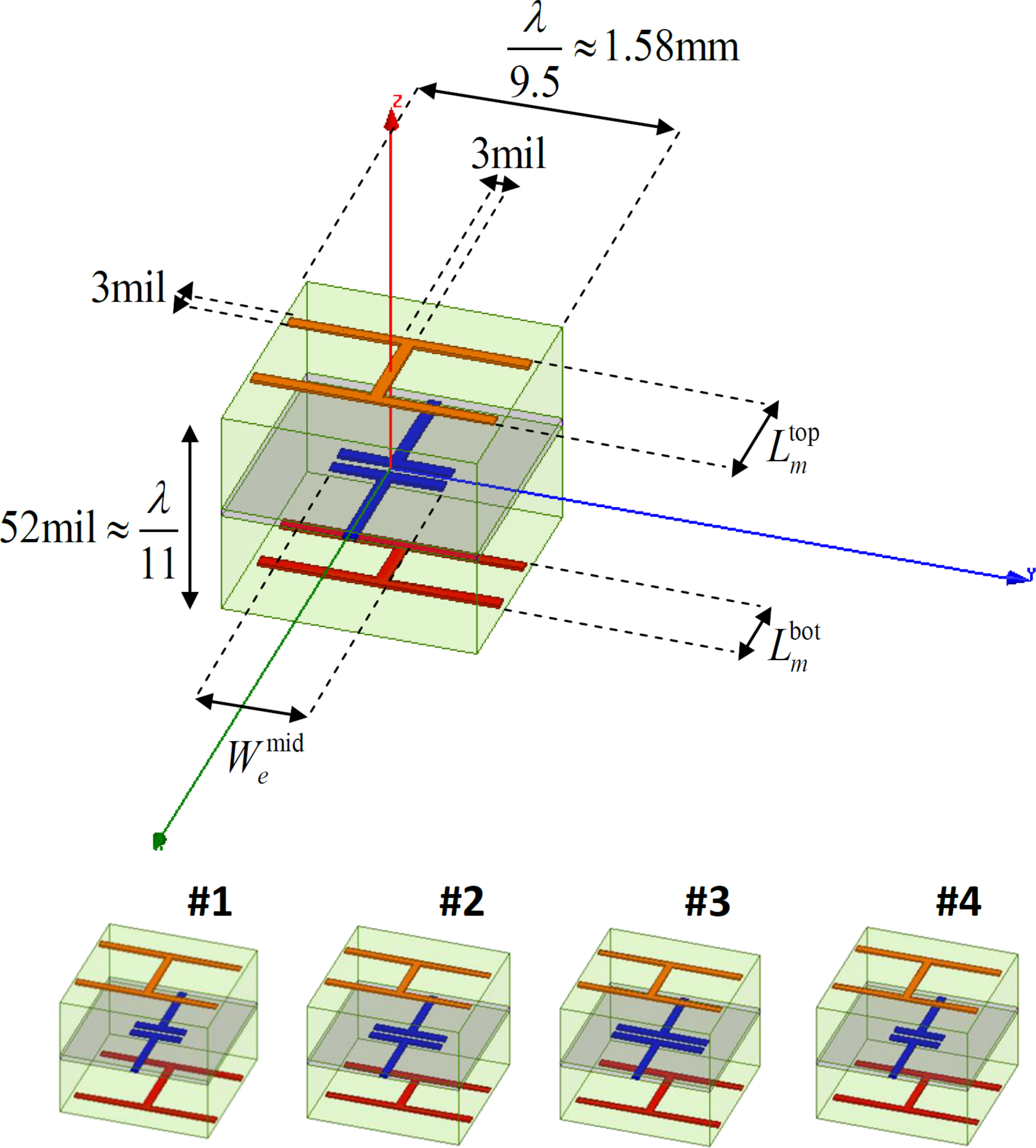}%
\caption{\label{fig:physical_meta_atom}Physical configuration of an asymmetric "spider" unit cell implementing an omega-type bianisotropic meta-atom. The three O-BMS degrees of freedom are tuned via the bottom dogbone arm length $L_m^\mathrm{bot}$, middle capacitor width $W_e^\mathrm{mid}$, and top dogbone arm length $L_m^\mathrm{top}$. The capacitor gap size and copper trace widths are fixed to $3\mathrm{mil}\approx76\mathrm{\mu m}$. The specific unit cells characterized in Table \ref{tab:meta_atoms} are presented at the bottom panels.}
}
\end{figure}

\textcolor{black}{
To demonstrate the viability of our microscopic design concept (Subsection \ref{subsec:micro_design}), we provide an example for a possible physical realization of representative omega-type bianisotropic meta-atoms for the application of engineered reflectionless refraction (Subsection \ref{subsec:refraction}). The unit cells are an asymmetric version of the spider unit cells presented in \cite{Epstein2016}, featuring three copper layers (Fig. \ref{fig:physical_meta_atom}) defined on two 25mil Rogers RT6010/Duroid laminates (permittivity $\epsilon=13.3\epsilon_0$ and loss tangent $\tan\delta=0.0023$), bonded using a 2mil-thick Rogers 2929 bondply (permittivity $\epsilon=2.94\epsilon_0$ and loss tangent $\tan\delta=0.003$). The three reactive sheets of Fig. \ref{fig:circuit_model} correspond to the bottom dogbone, middle loaded dipole, and the top dogbone; the reactance values are controlled by the dogbone arm lengths $L_m^\mathrm{bot}$ and $L_m^\mathrm{top}$, and the capacitor width $W_e^\mathrm{mid}$. At the design frequency $f=20\mathrm{GHz}$ the lateral dimensions of the unit cells are $\frac{\lambda}{9.5}\times\frac{\lambda}{9.5}\approx 1.58\mathrm{mm}\times1.58\mathrm{mm}$, and the overall metasurface thickness corresponds to $52\mathrm{mil}=1.32\mathrm{mm}\approx \frac{\lambda}{11}$. The copper traces are $18\mathrm{\mu m}$ thick, corresponding to a cladding of $1/2\mathrm{oz.}$, simulated using the standard bulk conductivity value of $\sigma=58\times 10^6 \mathrm{S/m}$.
}

\begin{table}[!t]
\centering
\textcolor{black}{
\begin{threeparttable}[b]
\renewcommand{\arraystretch}{1.3}
\caption{Realized O-BMS meta-atoms for engineered reflectionless refraction (Subsection \ref{subsec:refraction}), corresponding to Fig. \ref{fig:physical_meta_atom}}
\label{tab:meta_atoms}
\centering
\footnotesize{
\begin{tabular}{l|c c c c}
\hline \hline
\begin{tabular}{l} Cell \end{tabular} &  $\#1$ & $\#2$ & $\#3$ & $\#4$  \\
\hline \hline \\[-1.3em]	
	 \begin{tabular}{l} $L_m^\mathrm{bot}$ [3mil] \end{tabular}
	 & $12.00$ & $8.60$ & $10.60$ & $9.90$	 
	 	 	\\	\hline 
	 \begin{tabular}{l} $W_e^\mathrm{mid}$ [3mil] \end{tabular}
	 & $7.00$ & $9.70$ & $12.40$ & $6.60$	 
	 	 	 	 \\	\hline 
	 \begin{tabular}{l} $L_m^\mathrm{top}$ [3mil] \end{tabular}
	 & $11.00$ & $9.00$ & $9.37$ & $10.64$	 
	 	 	 	 \\	\hline 
	 \begin{tabular}{l} $X_{se} [\eta]$ (design) \end{tabular}
	 & $1.27$ & $-2.55$ & $-0.45$ & $0.12$	 
	 	 	 	 \\	\hline 
	 \begin{tabular}{l} $X_{se} [\eta]$ (realized) \end{tabular}
	 & $1.15$ & $-2.54$ & $-0.47$ & $0.14$	 
	 	 	 	 \\	\hline 
	 \begin{tabular}{l} $B_{sm} [1/\eta]$ (design) \end{tabular}
	 & $0.40$ & $-0.80$ & $-0.14$ & $0.036$	 
	 	 	 	 \\	\hline 
	 \begin{tabular}{l} $B_{sm} [1/\eta]$ (realized) \end{tabular}
	 & $0.52$ & $-0.88$ & $-0.14$ & $0.036$	 
	 	 	 	 \\	\hline 
	 \begin{tabular}{l} $K_{em} [-]$ (design) \end{tabular}
	 & $0.13$ & $0.68$ & $-0.10$ & $-0.14$	 
	 	 	 	 \\	\hline 
	 \begin{tabular}{l} $K_{em} [-]$ (realized) \end{tabular}
	 & $0.27$ & $0.72$ & $-0.09$ & $-0.13$	 
	 	 	 	 \\	\hline 
	 \begin{tabular}{l} $\angle G_{21}$ (design) \end{tabular}
	 & $74^\circ$ & $-52^\circ$ & $-124^\circ$ & $164^\circ$	 
	 	 	 	 \\	\hline 
	 \begin{tabular}{l} $\angle G_{21}$ (realized) \end{tabular}
	 & $75^\circ$ & $-49^\circ$ & $-123^\circ$ & $162^\circ$	 
	 	 	 	 \\	\hline 
	 \begin{tabular}{l} $\left|G_{21}\right|$ [dB] (realized) \end{tabular}
	 & $-1.78$ & $-0.73$ & $-0.58$ & $-0.89$	 
	 	 	 	 \\	\hline 
	 \begin{tabular}{l} $\left|G_{11}\right|$ [dB] (realized) \end{tabular}
	 & $-11.1$ & $-30.2$ & $-28.6$ & $-22.7$	 
	 	 	 	 \\
\hline \hline
\end{tabular}}
\end{threeparttable}}
\end{table}

\textcolor{black}{The four representative meta-atoms presented at the bottom of Fig. \ref{fig:physical_meta_atom} were designed to implement unit cells of the reflectionless refracting O-BMS designed in Subsection \ref{subsec:refraction} to couple a normally-incident plane wave to a transmitted plane wave propagating at $\theta_\mathrm{out}=71.81^\circ$, incurring a phase shift of $\xi_\mathrm{out}=70^\circ$. The physical structure is simulated in a periodic environment (\textit{cf.} \cite{Epstein2015_3,Pfeiffer2014_3}); by reversing \eqref{equ:OBMS_Z_matrix_explicit}, the resulting impedance matrix is used to characterize the corresponding local electric, magnetic, and magnetoelectric response.
As the metasurface local constituents $X_{se}=\Im\left\{Z_{se}\right\}$, $B_{sm}=\Im\left\{Y_{sm}\right\}$, and $K_{em}$ of \eqref{equ:OBMS_specifications_refraction} can be conveniently expressed as generalized scattering matrix parameters \eqref{equ:G_matrix_refraction}, the meta-atom geometrical degrees of freedom were swept to maximize the (generalized) transmission coefficient $\left|G_{21}\right|$ for certain (generalized) transmission phase values $\angle G_{21}$. Specifically, the cells numbered $\# 1$ to $\# 4$ were designed to introduce phases of $\angle G_{21}=74^\circ$, $\angle G_{21}=-52^\circ$, $\angle G_{21}=-124^\circ$, and $\angle G_{21}=164^\circ$, respectively, while retaining impedance matching to the respective incident and transmitted plane waves. As discussed towards the end of Subsection \ref{subsec:micro_design}, the sheet reactance values assessed via \eqref{equ:OBMS_impedance_sheets} for this meta-atom configuration ($\epsilon_\mathrm{sub}=13.3\epsilon_0$, $t=25\mathrm{mil}$) were used to choose the proper parameter range for an efficient sweep (similar to \cite{Pfeiffer2014_3}).
}

\textcolor{black}{The geometrical parameters and corresponding bianisotropic characteristics for these four unit cells are presented in Table \ref{tab:meta_atoms}. As can be observed therein, the O-BMS constituents extracted from the simulated impedance matrix by reversing \eqref{equ:OBMS_Z_matrix_explicit} agree well with the desirable ones. Furthermore, most unit cells are well matched to both wave impedances, establishing a low (generalized) reflection coefficient and a high (generalized) transmission coefficient, with very good correspondence between the required and realized transmission phase. Cell $\# 1$ shows slightly reduced performance, due to pronounced losses; as has been demonstrated in \cite{Pfeiffer2014_3} for chiral bianisotropic meta-atoms, we believe the effect of losses can be mitigated by introducing additional degrees of freedom, e.g. using a four-layer impedance sheet structure. Although a detailed design is beyond the scope of this paper, these results demonstrate that general O-BMS meta-atoms can indeed be realized with three metallic layers supported by standard microwave laminates and bondply, allowing fabrication via common PCB manufacturing technology.
}


\section*{Acknowledgment}
Financial support from the Natural Sciences and Engineering Research Council of
Canada (NSERC) is gratefully acknowledged. A.E. gratefully acknowledges the support of
the Lyon Sachs Postdoctoral Fellowship Foundation as well as the Andrew and Erna
Finci Viterbi Postdoctoral Fellowship Foundation of the Technion - Israel Institute of
Technology, Haifa, Israel.

\ifCLASSOPTIONcaptionsoff
  \newpage
\fi



%

%

\begin{IEEEbiography}[{\includegraphics[width=1in,height=1.25in,clip,keepaspectratio]{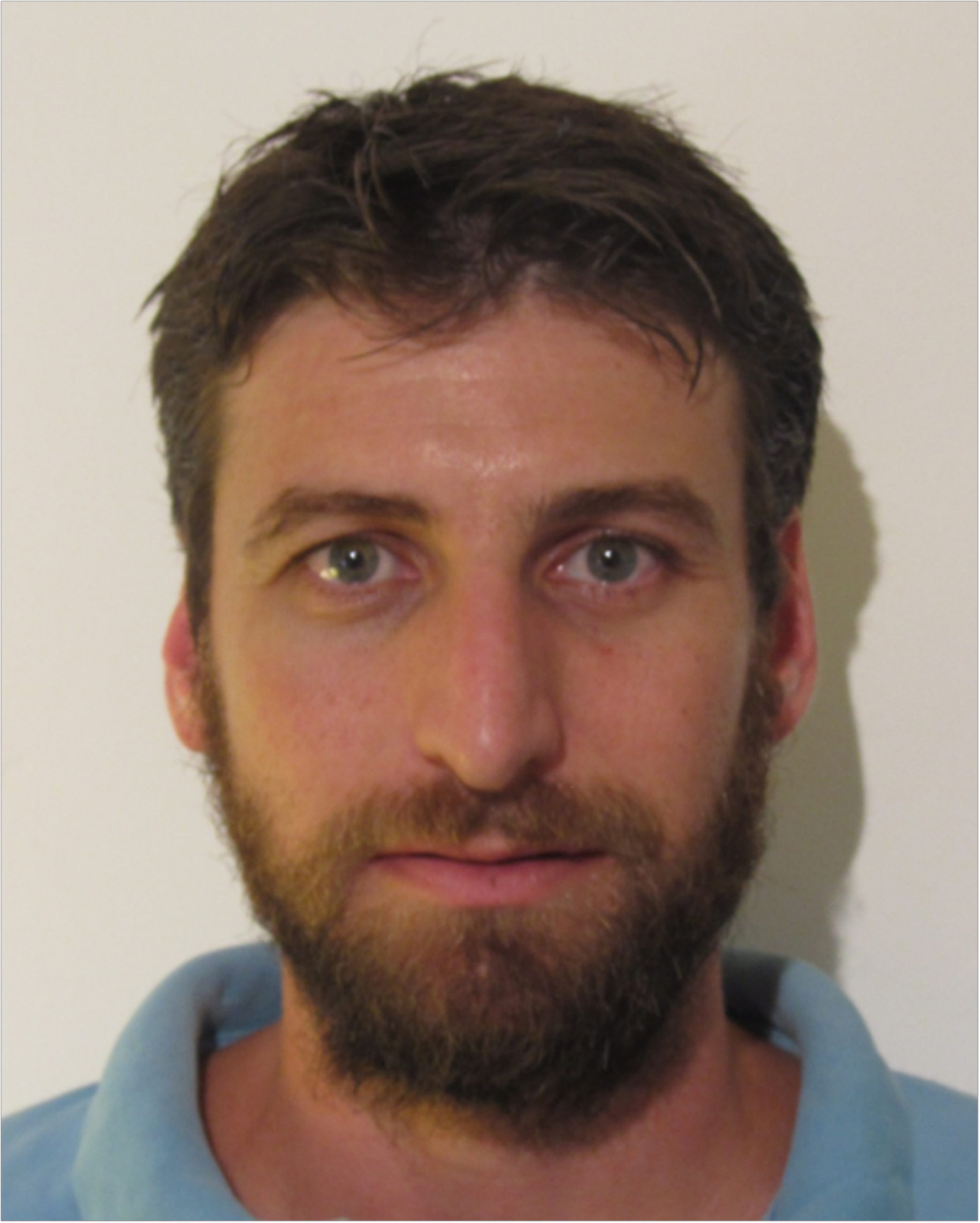}}]{Ariel Epstein}
(S'12-M'14) received the B.A. degree in computer science from the Open University of Israel, Raanana, the B.A. degree in physics, and the B.Sc. and Ph.D. degrees in electrical engineering from the Technion - Israel Institute of Technology, Haifa, Israel, in 2000, 2003, and 2013, respectively. Currently, he is a Lyon Sachs Postdoctoral Fellow in the Department of Electrical and Computer Engineering in the University of Toronto, Toronto, ON, Canada. His current research interests include utilization of electromagnetic theory, with an emphasis on analytical techniques, for the development of novel metasurface-based antenna and microwave devices, and investigation of new physical effects.

Dr. Epstein was the recipient of the Young Scientist Best Paper Award in the URSI Commission B International Symposium on Electromagnetic Theory (EMTS2013), held in Hiroshima, Japan, in May 2013, as well as the Best Poster Award at the 11th International Symposium on Functional $\pi$-electron Systems (F$\pi$-11), held in Arcachon, France, in June 2013.
\end{IEEEbiography}

\begin{IEEEbiography}[{\includegraphics[width=1in,height=1.25in,clip,keepaspectratio]{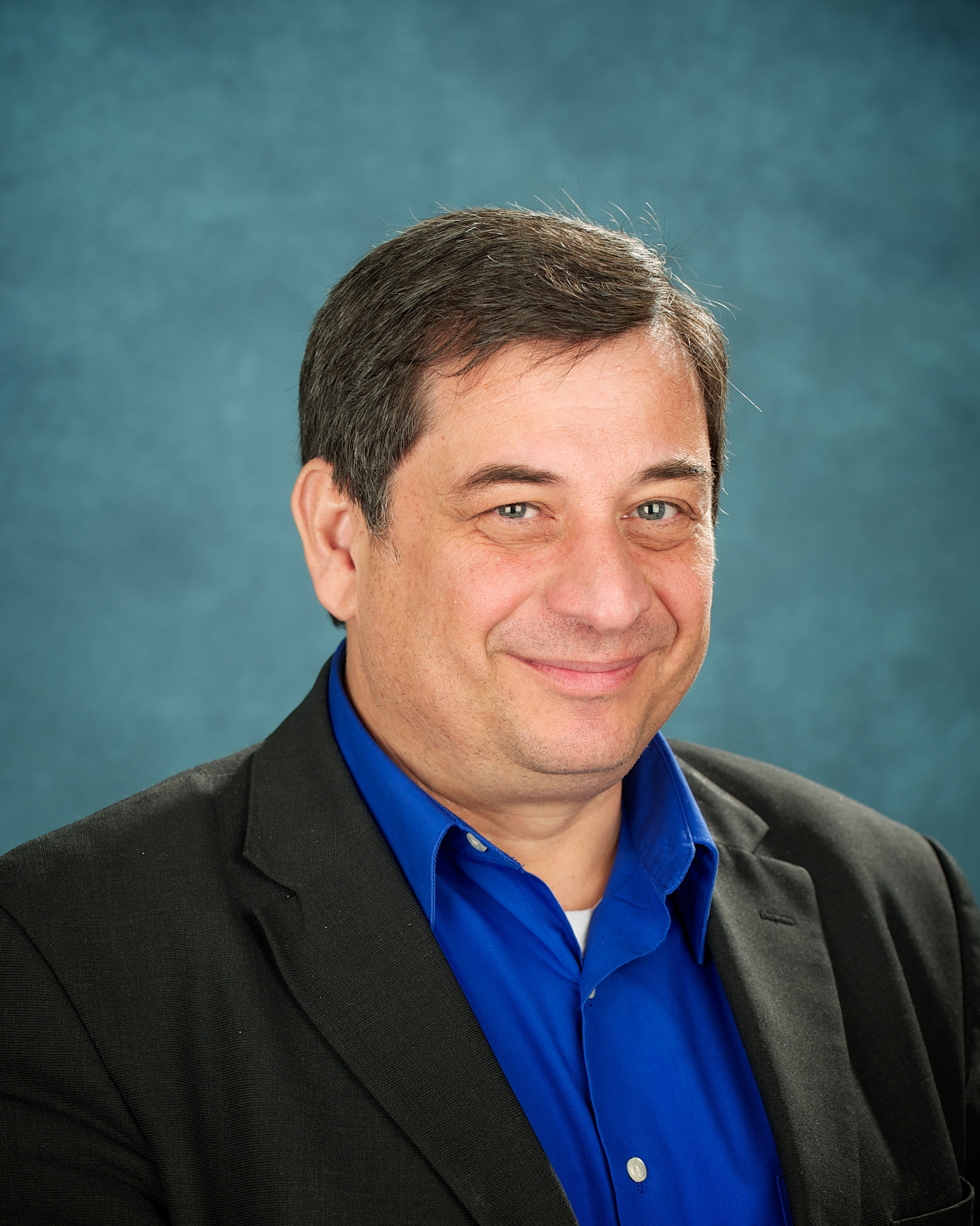}}]{George V. Eleftheriades}
(S'86-M'88-SM'02-F'06) earned the M.S.E.E. and Ph.D. degrees in electrical engineering from the University of Michigan, Ann Arbor, MI, USA, in 1989 and 1993, respectively. From 1994 to 1997, he was with the Swiss Federal Institute of Technology, Lausanne, Switzerland. Currently, he is a Professor in the Department of Electrical and Computer Engineering at the University of Toronto, ON, Canada, where he holds the Canada Research/Velma M. Rogers Graham Chair in Nano- and Micro-Structured Electromagnetic Materials. He is a recognized international authority and pioneer in the area of metamaterials. These are man-made materials which have electromagnetic properties not found in nature. He introduced a method for synthesizing metamaterials using loaded transmission lines. Together with his graduate students, he provided the first experimental evidence of imaging beyond the diffraction limit and pioneered several novel antennas and microwave components using these transmission-line based metamaterials. His research has impacted the field by demonstrating the unique electromagnetic properties of metamaterials; used in lenses, antennas, and other microwave and optical components to drive innovation in fields such as satellite communications, defence, medical imaging, microscopy, automotive radar, and wireless telecommunications. Presently, he is leading a group  of 12 graduate students and researchers in the areas of electromagnetic and optical metamaterials, and  metasurfaces, antennas and components for broadband wireless communications, novel antenna beam-steering techniques, far-field super-resolution imaging, radars, plasmonic and nanoscale optical components, and fundamental electromagnetic theory. 

Prof. Eleftheriades served as an Associate Editor for the IEEE TRANSACTIONS ON ANTENNAS AND PROPAGATION (AP). He also served as a member of the IEEE AP-Society administrative committee (AdCom) from 2007 to 2012 and was an IEEE AP-S Distinguished Lecturer from 2004 to 2009. He served as the General Chair of the 2010 IEEE International Symposium on Antennas and Propagation held in Toronto, ON, Canada. Papers that he co-authored have received numerous awards such as the 2009 Best Paper Award from the IEEE MICROWAVE AND WIRELESS PROPAGATION LETTERS,  twice the R. W. P. King Best Paper Award from the IEEE TRANSACTIONS ON ANTENNAS AND PROPAGATION (2008 and 2012), and the 2014 Piergiorgio Uslenghi Best Paper Award from the IEEE ANTENNAS AND WIRELESS PROPAGATION LETTERS. His work has been cited more than 12,000 times. He received the Ontario Premier’s Research Excellence Award and the University of Toronto’s Gordon Slemon Award, both in 2001. In 2004 he received an E.W.R. Steacie Fellowship from the Natural Sciences and Engineering Research Council of Canada. He is the recipient of the 2008 IEEE Kiyo Tomiyasu Technical Field Award and the 2015 IEEE John Kraus Antenna Award. In 2009, he was elected a Fellow of the Royal Society of Canada.
\end{IEEEbiography}

%
%




\end{document}